\documentclass[english]{lipics}

\usepackage{amsmath,amssymb,stmaryrd,epsf,xcolor,listings}

\newcommand\fa{\forall}
\newcommand\ex{\exists}
\newcommand\lra{\longrightarrow}
\newcommand\ra{\rightarrow}
\newcommand\T{{\cal T}}
\newcommand\eqbg{\equiv_{\beta\Gamma}}
\newcommand\redbg{\rightarrow_{\beta\Gamma}}

\DeclareMathOperator{\ato}{\triangleright}

\newcommand\lst[1]{{\rm \lstinline{#1}}}

\newbox\tempa
\newbox\tempb
\newdimen\tempc
\def\mud#1{\hfil $\displaystyle{\mathstrut #1}$\hfil}
\def\rig#1{\hfil $\displaystyle{#1}$}
\def\irulehelp#1#2#3{\setbox\tempa=\hbox{$\displaystyle{\mathstrut #2}$}%
                        \setbox\tempb=\vbox{\halign{##\cr
        \mud{#1}\cr
        \noalign{\vskip\the\lineskip}
        \noalign{\hrule height 0pt}
        \rig{\vbox to 0pt{\vss\hbox to 0pt{${\; #3}$\hss}\vss}}\cr
        \noalign{\hrule}
        \noalign{\vskip\the\lineskip}

        \mud{\copy\tempa}\cr}}
                      \tempc=\wd\tempb
                      \advance\tempc by \wd\tempa
                      \divide\tempc by 2 }
\def\irule#1#2#3{{\irulehelp{#1}{#2}{#3}
                     \hbox to \wd\tempa{\hss \box\tempb \hss}}}

\begin{document}
\title{
\textsc{Dedukti}: a Logical Framework based on the
$\lambda \Pi$-Calculus Modulo Theory}
\author[1]{Ali Assaf}
\author[2]{Guillaume Burel}
\author[3]{Rapha\"el Cauderlier}
\author[4]{David Delahaye}
\author[5]{Gilles Dowek}
\author[2]{Catherine Dubois}
\author[6]{Fr\'ed\'eric Gilbert}
\author[7]{Pierre Halmagrand}
\author[8]{Olivier Hermant}
\author[8]{Ronan Saillard}
\authorrunning{Deducteam}

\affil[1]{Inria and \'Ecole polytechnique, {\tt ali.assaf@inria.fr}}

\affil[2]{ENSIIE, {\tt \{guillaume.burel,catherine.dubois\}@ensiie.fr}}

\affil[3]{IRIF, University Paris Diderot, and CNRS,
  {\tt raphael.cauderlier@irif.fr}}

\affil[4]{CNAM and Université de Montpellier,
{\tt david.delahaye@umontpellier.fr}}

\affil[5]{Inria and \'Ecole Normale de Supérieure de Cachan,
{\tt gilles.dowek@ens-cachan.fr}}

\affil[6]{\'Ecole des Ponts, Inria, and CEA,
{\tt frederic.gilbert@inria.fr}}

\affil[7]{CNAM and Inria,
  {\tt pierre.halmagrand@inria.fr}}

\affil[8]{MINES ParisTech,
  {\tt \{olivier.hermant,ronan.saillard\}@mines-paristech.fr}}

\date{}
\maketitle
\thispagestyle{empty}

\begin{abstract}
\textsc{Dedukti} is a Logical Framework based on the $\lambda
\Pi$-Calculus Modulo Theory. We show that many theories can be
expressed in \textsc{Dedukti}: constructive and classical predicate
logic, Simple type theory, programming languages, Pure type systems,
the Calculus of inductive constructions with universes, etc. and that
permits to used it to check large libraries of proofs developed in
other proof systems: \textsc{Zenon}, \textsc{iProver},
\textsc{FoCaLiZe}, \textsc{HOL Light}, and \textsc{Matita}.
\end{abstract}

\section{Introduction}
\label{sec:intro}

Defining a theory, such as arithmetic, geometry, or set theory, in
predicate logic \cite{HilbertAckermann} just requires to choose
function and predicate symbols and axioms expressing the meaning of
these symbols. Using this way a single logical framework to define
all these theories has many advantages.

First, it requires less efforts, as the logical connectives, $\wedge$,
$\vee$, $\fa$, etc. and their associated deduction rules are defined,
in the framework, once and for all and need not be redefined for each
theory. Similarly, the notions of proof, model, etc. are defined once and
for all. And general theorems, such as the soundness and the
completeness theorems, can be proved once and for all.

Another advantage of using such a logical framework is that this
induces a partial order between theories. For instance,
Zermelo-Fraenkel set theory with the axiom of choice (ZFC) is an
extension of Zermelo-Fraenkel set theory (ZF), as it contains the same
axioms, plus the axiom of choice. It is thus obvious that any theorem
of ZF is provable in ZFC, and for each theorem of ZFC, we can ask the
question of its provability in ZF. Several theorems of ZFC that are
provable in ZF have been identified, and these theorems can be used
in extensions of ZF that are inconsistent with the axiom of choice.

Finally, using such a common framework permits to combine, in a proof,
lemmas proved in different theories: if $\T$ is a theory expressed in
a language ${\cal L}$ and $\T'$ a theory expressed in a language
${\cal L}'$, if $A$ is expressed in ${\cal L} \cap {\cal L}'$, $A
\Rightarrow B$ is provable in $\T$, and $A$ is provable in $\T'$, then
$B$ is provable in $\T \cup \T'$.

Despite these advantages, several logical systems have been defined,
not as theories in predicate logic, but as independent systems: Simple
type theory \cite{Church40,Andrews86}, also known as Higher-order
logic, is defined as an independent system---although it is also
possible to express it as a theory in predicate logic. Similarly,
Intuitionistic type theory \cite{MartinLof}, the Calculus of
constructions \cite{CoquandHuet}, the Calculus of inductive
constructions \cite{Paulin-Mohring}, etc. are defined as independent
systems.  As a consequence, it is difficult to reuse a formal proof
developed in an automated or interactive theorem prover based on one
of these formalisms in another, without redeveloping it. It is also
difficult to combine lemmas proved in different systems: the realm of
formal proofs is today a tower of Babel, just like the realm of
theories was, before the design of predicate logic.

The reason why these formalisms have not been defined as theories in
predicate logic is that predicate logic, as a logical framework, has
several limitations, that make it difficult to express some logical
systems.
\begin{enumerate}
\item\label{problem-bound-variables} Predicate logic does not allow
  the use of bound variables, except those bound by the quantifiers
  $\fa$ and $\ex$. For instance, it is not possible to define, in
  predicate logic, a unary function symbol $\mapsto$ that would bind
  a variable in its argument.

\item\label{problem-propositions-as-types} Predicate logic ignores the
  propositions-as-types principle, according to which a proof $\pi$ of
  a proposition $A$ is a term of type $A$.

\item\label{problem-computation} Predicate logic ignores the
  difference between deduction and computation. For example, when
  Peano arithmetic is presented in predicate logic, there is no
  natural way to compute the term $2 \times 2$ into $4$. To prove the
  theorem $2 \times 2 = 4$, several deduction steps need to be used
  while a simple computation would have sufficed \cite{Poincare}.

\item\label{problem-cut} Unlike the notions of proof and model, it is
  not possible to define, once and for all, the notion of cut in
  predicate logic and to apply it to all theories expressed in
  predicate logic: a specific notion of cut must be defined for each
  theory.

\item\label{problem-classical} Predicate logic is classical and not
  constructive. Constructive theories must be defined in another
  logical framework: constructive predicate logic.
\end{enumerate}

This has justified the development of other logical frameworks, that
address some of these problems.  Problem \ref{problem-bound-variables}
has been solved in extensions of predicate logic such as
\textsc{$\lambda$-Prolog} \cite{MillerNadathur} and \textsc{Isabelle}
\cite{Paulson}.
Problems \ref{problem-bound-variables} and
\ref{problem-propositions-as-types} have been solved in an extension
of predicate logic, called the Logical framework \cite{HHP}, also
known as the $\lambda \Pi$-calculus, and the $\lambda$-calculus with
dependent types.  Problems \ref{problem-computation} and
\ref{problem-cut} have been solved in an extension of predicate logic,
called Deduction modulo theory \cite{DHK}.  Combining the $\lambda
\Pi$-calculus and Deduction modulo theory yields the $\lambda
\Pi$-calculus modulo theory \cite{CousineauDowek}, a variant of
Martin-L\"of's logical framework \cite{NPS}, which solves problems
\ref{problem-bound-variables}, \ref{problem-propositions-as-types},
\ref{problem-computation}, and \ref{problem-cut}.

Problem \ref{problem-classical} has been addressed
\cite{Doweknotnot,HermantGilbert,Prawitz} and will be discussed in
Section \ref{sec:classical}.

The expressivity of the $\lambda \Pi$-calculus modulo theory has
already been discussed \cite{CousineauDowek}. We have shown
that all functional Pure type systems \cite{Barendregt1992}, in
particular the Calculus of constructions, could be expressed in this
framework.

Our goal, in this paper, is twofold: first, we want to go further and
show that other systems can be expressed in the $\lambda \Pi$-calculus
modulo theory, in particular classical systems, logical systems
containing a programming language as a subsystem,
Simple type theory, and extensions of the Calculus
of constructions with universes and inductive types.

Second, we want to demonstrate this expressivity, not just with
adequacy theorems, but also by showing that large libraries of formal
proofs coming from automated and interactive theorem provers can be
translated to and checked in \textsc{Dedukti}, our implementation of
the $\lambda \Pi$-calculus modulo theory. To do so, we translate
to \textsc{Dedukti} proofs developed in various systems: the automated
theorem proving systems \textsc{Zenon modulo}, \textsc{iProverModulo},
the system containing a programming language as a subsystem
\textsc{FoCaLiZe}, and the interactive theorem provers \textsc{HOL
  Light} and \textsc{Matita}.  Expressing formal proofs, coming from
different systems, in a single logical framework is a first step
towards reverse engineering these proofs, that is precisely analyzing
the theories in which these theorems can be proved. It is also a first
step towards interoperability, that is building proofs assembling
lemmas developed in different systems. Finally, it is also a
way to audit formal proofs develop in these systems by checking them
in a different, independent, system.

We show this way that the logical framework approach scales up and
that it is mature enough to start imagining a single distributed
library of formal proofs expressed in different theories, developed in
different systems, but formulated in a single framework.

We present here a synthesis of the state of the art about the
$\lambda \Pi$-calculus modulo theory, its implementation
\textsc{Dedukti}, and the expression of theories in this
framework. In Sections \ref{sec:constructive}, \ref{sec:classical},
\ref{sec:simple-type-theory}, \ref{sec:programming-languages}, and
\ref{sec:cicw}, we shall present a number of examples of theories that
have been expressed in \textsc{Dedukti} and of libraries of proofs
expressed in these theories. Before that, we shall present the
$\lambda \Pi$-calculus modulo theory, in Section~\ref{sec:lpm}, and
the \textsc{Dedukti} system, in Section~\ref{sec:dk}.

\section{The  $\lambda \Pi$-calculus modulo theory}
\label{sec:lpm}

We present the $\lambda \Pi$-calculus modulo theory in two steps. We
first present, in Section \ref{subsec:lpm}, a very general system, not
caring about the decidability of type-checking. Then, in Section
\ref{subsec:effective}, we discuss restrictions of this system that
enforce the decidability of type-checking.

\subsection{The  $\lambda \Pi$-calculus}
\label{subsec:lp}

The $\lambda \Pi$-calculus is an extension of the simply typed
$\lambda$-calculus with dependent types. In this calculus, types are
just terms of a particular type $Type$, some symbols have the type $A
\ra Type$, and can be applied to a term of type $A$ to build a type
depending on this term. A type $Kind$ is introduced to type the terms
$Type$, $A \ra Type$, etc. and the arrow $A \ra B$ is extended to a
dependent product $\Pi x:A~B$. Finally, besides the usual typing
judgements, the typing rules use another kind of judgement expressing
the well-formedness of a context. The typing rules are the following.

\noindent
Well-formedness of the empty context
$$\irule{}
        {[~]~\mbox{well-formed}}
        {}$$
Declaration of a type or type family variable
$$\irule{\Gamma \vdash A:Kind}
        {\Gamma, x:A~\mbox{well-formed}}
        {}$$
Declaration of an object variable
$$\irule{\Gamma \vdash A:Type}
        {\Gamma, x:A~\mbox{well-formed}}
        {}$$
$Type$
$$\irule{\Gamma~\mbox{well-formed}}
        {\Gamma \vdash Type:Kind}
        {}$$
Variable
$$\irule{\Gamma~\mbox{well-formed}}
        {\Gamma \vdash x:A}
        {x:A \in \Gamma}$$
Product (for kinds)
$$\irule{\Gamma \vdash A:Type~~~\Gamma, x:A \vdash B:Kind}
        {\Gamma \vdash \Pi x:A~B:Kind}
        {}$$
Product (for types)
$$\irule{\Gamma \vdash A:Type~~~\Gamma, x:A \vdash B:Type}
        {\Gamma \vdash \Pi x:A~B:Type}
        {}$$
Abstraction (for type families)
$$\irule{\Gamma \vdash A:Type~~~\Gamma, x:A \vdash B:Kind~~~
         \Gamma, x:A \vdash t:B}
        {\Gamma \vdash \lambda x:A~t:\Pi x:A~B}
        {}$$
Abstraction (for objects)
$$\irule{\Gamma \vdash A:Type~~~\Gamma, x:A \vdash B:Type~~~
         \Gamma, x:A \vdash t:B}
        {\Gamma \vdash \lambda x:A~t:\Pi x:A~B}
        {}$$
Application
$$\irule{\Gamma \vdash t:\Pi x:A~B~~~\Gamma \vdash t':A}
        {\Gamma \vdash (t~t'):(t'/x)B}
        {}$$
Conversion
$$\irule{\Gamma \vdash t:A~~~\Gamma \vdash A:Type~~~\Gamma \vdash B:Type}
        {\Gamma \vdash t:B}
        {A \equiv_\beta B}$$
$$\irule{\Gamma \vdash t:A~~~\Gamma \vdash A:Kind~~~\Gamma \vdash B:Kind}
        {\Gamma \vdash t:B}
        {A \equiv_\beta B}$$

\subsection{The $\lambda \Pi$-calculus modulo theory}
\label{subsec:lpm}

The $\lambda \Pi$-calculus modulo theory is an extension of the
$\lambda \Pi$-calculus. In the $\lambda \Pi$-calculus modulo theory,
the contexts contain variable declarations---like in the $\lambda
\Pi$-calculus---and also rewrite rules.  The conversion rule is
extended in order to take these rewrite rules into account.

\subsubsection{Local and global contexts}

There are two notions of context in the $\lambda \Pi$-calculus modulo
theory: {\em global} contexts and {\em local} ones. Global contexts
contain variable declarations and rewrite rules, but local ones
contain object variable declarations only, that is declaration of
variables whose type has type $Type$ and not $Kind$. A third kind of
judgement is introduced to express that $\Delta$ is a local context in
$\Gamma$, and two rules to derive such judgements.

\noindent
Locality of the empty context
$$\irule{\Gamma~\mbox{well-formed}}
        {\Gamma \vdash [~]~\mbox{local}}
        {}$$
Declaration of an object variable in a local context
$$\irule{\Gamma \vdash \Delta~\mbox{local}~~\Gamma, \Delta \vdash A : Type}
        {\Gamma \vdash \Delta, x : A~\mbox{local}}
        {}$$

The following lemma is proved with a simple induction on the
structure of the derivation of the judgment ``$\Delta$ local''.
\begin{lemma}
If $\Delta$ is local in $\Gamma$, then $\Gamma, \Delta$ is a
well-formed global context.
\end{lemma}

Thus, defining a local context is just a way to express that some end
segment of the context contains object variables only.

A rewrite rule in a global context $\Gamma$ is a pair of terms
$\langle l,r \rangle$, where $l$ is a term of the form $f~u_1~...~u_n$
for some variable $f$, together with a context $\Delta$ local in
$\Gamma$. We note it $l\lra^\Delta r$.

The notion of well-formed global context is extended to allow
the declaration of rewrite rules
$$\irule{\Gamma~\mbox{well-formed}~~\Gamma\vdash\Delta\text{ local}}
{\Gamma, l \lra^\Delta r~\mbox{well-formed}} {}$$ Note that, at this
stage, $l$ and $r$ can have different types or be not well-typed at
all. Conditions will be added in Section \ref{subsec:effective}.

\subsubsection{Conversion}

If $l \lra^\Delta r$ is a rewrite rule and $\sigma$ is a substitution binding
the variables of $\Delta$, we say that the term $\sigma l$ {\em rewrites} to
the term $\sigma r$.

If $\Gamma$ is a well-formed global context, then the rewriting
relation generated by $\beta$-reduction and $\Gamma$, noted $\redbg$,
is the smallest relation, closed by context, such that if $t$ rewrites
to $u$, for
some rule in $\Gamma$, or if $t$ $\beta$-reduces to $u$, then $t
\redbg u$.  The congruence $\eqbg$ is the
reflexive-symmetric-transitive closure of the relation $\redbg$.

The conversion rules of the $\lambda \Pi$-calculus modulo theory are
the same as the conversion rules of the $\lambda \Pi$-calculus except that
the congruence $\equiv_\beta$ is replaced by $\eqbg$
$$\irule{\Gamma \vdash A:Type~~\Gamma \vdash B:Type~~\Gamma \vdash t:A}
        {\Gamma \vdash t:B}
        {A \eqbg B}$$
$$\irule{\Gamma \vdash A:Kind~~\Gamma \vdash B:Kind~~\Gamma \vdash t:A}
        {\Gamma \vdash t:B}
        {A \eqbg B}$$

\subsubsection{Subject reduction}\label{subsec:lampi:sr}

Subject reduction expresses that reduction preserves
the type of a term.  This property does not hold in general in the
$\lambda \Pi$-calculus modulo theory. However it does if we have the two
following properties: {\em well-typedness} of rewrite rules and
{\em product compatibility}
\cite{Barbaneraetal}. Henceforth, unless stated otherwise,
$\Gamma \vdash t: T$ stands for ``the judgement $\Gamma \vdash t: T$
is derivable''.

A rewrite rule is {\em well-typed} if it is type preserving.

\begin{definition}\label{def:WellTypedRewriteRule}
	Let  $l \lra^{\Delta_0} r$ be a rewrite rule in the context $\Gamma_0$
	and $\Gamma$ be a well-formed extension of $\Gamma_0$.
	We say that $l \lra^{\Delta_0} r$ is {\em well-typed} for $\Gamma$
	if for any substitution $\sigma$ binding the variables of $\Delta_0$
	$$ \mbox{if } \Gamma \vdash \sigma l : T  \mbox{,
          then } \Gamma \vdash \sigma r :T$$
\end{definition}

\begin{definition}

A well-formed global context $\Gamma$ satisfies the {\em product
compatibility} property if for any products $\Pi x:A_1~B_1$ and
$\Pi x:A_2~B_2$ well-typed in $\Gamma$
$$\mbox{if } \Pi x:A_1~B_1 \eqbg \Pi x:A_2~B_2 \mbox{ then } A_1 \eqbg
A_2~\text{and}~B_1 \eqbg B_2 \mbox{.}$$
\end{definition}

Note that this property immediately follows from the confluence of
$\beta\Gamma$-reduction, but it is weaker.

From well-typedness of rewrite rules and product compatibility, we can prove
subject reduction.

\begin{lemma}[Subject Reduction, Theorem 2.6.22 in \cite{Saillard15}]
Let $\Gamma$ be a global context satisfying product compatibility
and whose rewrite rules are well-typed.
If $\Gamma \vdash t_1 : T$ and $t_1 \redbg t_2$
then $\Gamma \vdash t_2 : T$.
\end{lemma}

Using product compatibility we can also show that a term has, at most,
one type modulo conversion.

\begin{lemma}[Uniqueness of Types, Theorem 2.6.34 in \cite{Saillard15}]
Let $\Gamma$ be a global context satisfying the product compatibility property.
If $\Gamma \vdash t : T_1$ and $\Gamma \vdash t : T_2$
then $T_1 \eqbg T_2$.
\end{lemma}

\subsubsection{Rules and rule schemes}
\label{sec:infinite}

In predicate logic, some theories are defined with a finite set of
axioms, in which case these axioms can just be enumerated. Some
others, such as arithmetic or set theory, are defined with an infinite
but decidable set of axioms: an axiom scheme.  In a specific proof
however, only a finite number of such axioms can be used. Thus,
provability of a proposition $A$ in a theory $\T$ having axiom schemes
can either be defined as the fact that the infinite sequent $\T \vdash
A$ has a proof, or as the fact that there exists a finite subset
$\Gamma$ of $\T$, such that the finite sequent $\Gamma \vdash A$ has a
proof.

In the same way, some theories in the $\lambda \Pi$-calculus modulo
theory may be defined with an infinite set of symbols and an infinite
set of rewrite rules, recognized by an algorithm. In this case, as the
contexts are always finite in the $\lambda \Pi$-calculus modulo theory,
the derivability of a typing relation $t:A$ must be
defined as the existence of a finite context $\Gamma$, such that the
judgment $\Gamma \vdash t:A$ is derivable.

\subsection{Effective subsystems}
\label{subsec:effective}

In the calculus presented in Section \ref{subsec:lpm},
proof-checking is not always decidable because, when using the
conversion rule to reason modulo the congruence,
$$\irule{\Gamma \vdash A:s~~\Gamma \vdash B:s~~\Gamma \vdash t:A}
        {\Gamma \vdash t:B}
        {A \eqbg B}$$
the conversion sequence justifying $A \eqbg B$ is not recorded in the
conclusion.  Thus, we need to restrict to {\em effective} subsystems,
that is systems where the rewrite rules are restricted in such a way
that the congruence $\eqbg$ is decidable.  A natural subsystem is
obtained by restricting the rewrite rules to form, together with the
$\beta$-reduction rule, a terminating and confluent rewrite
system. This way, the congruence can be decided by checking
equality of normal forms.

As $\beta$-reduction does not terminate on untyped terms, termination
can hold only for typed ones. Therefore, the subject reduction
property is essential to prove the termination of
the calculus.  As we have seen in Section \ref{subsec:lampi:sr}, subject
reduction is a consequence of well-typedness of rewrite rules and
product compatibility, which is itself a consequence of
confluence. Thus confluence should be proved first, without assuming
termination or restricting ourselves to typed terms.

The rest of this section discusses effective conditions for confluence
and well-typedness.

\subsubsection{Higher-order rewriting}

Confluence of the rewrite relation together with $\beta$-reduction is not
general enough. Indeed, such a confluence result may easily be lost
when allowing $\lambda$-abstraction in the left-hand side of a rewrite
rule, a useful feature---see, for instance, Section \ref{universes}.

For example, let $\Delta = f:Real \ra Real$, and consider the rewrite
rule explaining how to differentiate the function $\lambda
x:Real~(\sin~(f~x))$
$$(D~(\lambda x:Real~(\sin~(f~x)))) \lra^\Delta (\times~(D~(\lambda
x:Real~(f~x)))~(\lambda x:Real~(\cos~(f~x))))$$
where $\times$ is the pointwise product of functions.
This rule overlaps with $\beta$-reduction and creates the critical peak
$$(D~(\lambda x~(\sin~x)))
\longleftarrow
(D~(\lambda x~(\sin~((\lambda y~y)~x))))
\lra
(\times~(D~(\lambda x~((\lambda y~y)~x)))~(\lambda x~(\cos~((\lambda y~y)~x))))$$
that cannot be joined. This rewrite system with $\beta$-reduction
is therefore not confluent.

Still, $\eqbg$ is decidable, as some form of confluence holds. For
this, we need a richer rewrite relation: rewriting modulo
$\beta$-equivalence, also known as higher-order rewriting.

Higher-order rewriting requires higher-order matching: the term $t$
matches the term $l$ if there is a substitution
$\sigma$ such that $\sigma l$ is $\beta$-convertible to
$t$. Higher-order matching is undecidable in general~
\cite{Dowek91,DBLP:journals/igpl/Loader03}, but it is decidable if
we restrict to Miller's patterns~\cite{Miller91alogic}.  Thus, we
restrict the left-hand sides of the rewrite rules to be patterns.

\begin{definition}
A $\beta$-normal term $f~u_1~...~u_n$ is a pattern with respect to
$\Delta$ if $f$ is a variable not declared in $\Delta$, and the
variables declared in $\Delta$ are applied to pairwise distinct
bound variables.
\end{definition}

For example, if $f$ is declared in $\Delta$, then the term
$D~(\lambda x~(\sin~(f~x)))$ is a pattern, but neither $D~(\lambda
x~(f~(\sin~x)))$ nor $D~((\lambda x~(\sin~x))~y)$ are.

Now, if we consider the derivative rule as a higher-order rule, we have
$$D~(\lambda x~(\sin~x)) \lra \times~(D~(\lambda x~x))~(\lambda x~(\cos~x))$$
thus the peak above is now joinable, and, more generally, confluence
holds.

Given a pattern $l$ and a term $t$ there are different substitutions
$\sigma$ such that $\sigma l \eqbg t$ and some of them are not be well-typed.
To get a precise
and type-safe definition of higher-order-rewriting, we use the
concepts of higher-order rewrite system~\cite{Nipkow-LICS-91}. Remark, however, that
our notion of higher-order rewriting does not require the identification of $\eta$-equivalent terms
as for higher-order rewrite systems.  See
\cite{RSai15} for more details. This also gives us powerful confluence
criteria such as confluence for development-closed
systems~\cite{DBLP:conf/hoa/Oostrom95}.

The reason why we are interested in the confluence of this extended
higher-order rewrite relation is that, when combined with
$\beta$-reduction, first- and higher-order rewriting yield the same
congruence. Therefore, if higher-order rewriting is confluent and
terminating, $\eqbg$ is decidable.

\subsubsection{Typing rewrite rules}
\label{subsec:typingrewriterules}

As we want reduction to preserve typing, we must check that the
rewrite rules are well-typed in the sense of Definition
\ref{def:WellTypedRewriteRule}. To do so, we introduce two new
judgments: $\Gamma \vdash l \lra^\Delta r$ expressing that the rule $l
\lra^\Delta r$ is well-typed in $\Gamma$ and {\em $\Gamma$ strongly
  well-formed}, expressing that all the rules declared in $\Gamma$ are
well-typed.  This judgment is defined by the rules
$$\irule{}
        {[~]~\mbox{strongly well-formed}}
        {}$$
$$\irule{\Gamma~\mbox{strongly well-formed}~~~\Gamma \vdash A:Type}
        {\Gamma, x:A~\mbox{strongly well-formed}}
        {}$$
$$\irule{\Gamma~\mbox{strongly well-formed}~~~\Gamma \vdash A:Kind}
        {\Gamma, x:A~\mbox{strongly well-formed}}
        {}$$
$$\irule{\Gamma~\text{strongly well-formed}~~~\Gamma \vdash l\lra^\Delta r}
        {\Gamma, l\lra^\Delta r \text{ strongly well-formed}}
        {}$$

Defining an effective rule for the judgment $\Gamma \vdash l
\lra^\Delta r$, that fits Definition~\ref{def:WellTypedRewriteRule},
is more difficult. This definition  quantifies over all possible
substitutions, so is not effective.
Thus, we need to under-approximate this notion and find rules that
express sufficient conditions for well-typedness.  A first attempt
is to require $l$ and $r$ to be well-typed and to have the same type
in $\Gamma, \Delta$
$$\irule{l\text{ is a pattern}~~~\Gamma,\Delta \vdash l : T
         ~~~\Gamma,\Delta \vdash r : T~~~dom(\Delta)\subseteq FV(l)}
        {\Gamma \vdash l \lra^\Delta r}
        {}$$
But this condition is too restrictive as shown by the following example.

Consider a context containing a variable $A$ of type $Type$, a
variable $Vector$ of type $nat \ra Type$, a variable $Nil$ of type
$(Vector~0)$, and a variable $Cons$ of type $\Pi n:nat~(A \ra (Vector~n)
\ra (Vector~(S~n)))$.  Assume that we want to define the $Tail$
function of type $\Pi n:nat~((Vector~(S~n)) \ra (Vector~n))$.

A first possibility is to use the non left-linear rule
$$(Tail~n~(Cons~n~a~l)) \lra l$$
but proving confluence of a non-linear rewrite system, not assuming
termination---as underlined above, confluence comes first---, is
difficult. Thus, we might prefer the linear
rule $$(Tail~n~(Cons~m~a~l)) \lra l$$ This rule is well-typed in the
sense of Definition~\ref{def:WellTypedRewriteRule}, but its left-hand
side is not well-typed: $(Cons~m~a~l)$ has type $(Vector~(S~m))$ and
$(Tail~n)$ expects a term of type $(Vector~(S~n))$. So we need a
weaker condition, in order to accept this rule \cite{Blanqui05,Saillard15}.

\noindent Typing the term $\sigma (Tail~n~(Cons~m~a~l))$ generates the
constraint
$$(Vector~(S~\sigma n)) \equiv (Vector~(S~\sigma m))$$
Using the injectivity of the functions $Vector$ and $S$, this constraint
boils down to
$$\sigma n \equiv \sigma m$$
so all the substitutions $\sigma$, such that $\sigma (Tail~n~(Cons~m~a~l))$
is well-typed, have the form $\eta \tau$ where $\eta$ is an arbitrary
substitution and $\tau = p/m, p/n$. The substitution $\tau$ is the most
general typing substitution for $l$.
Both the terms $\tau (Tail~n~(Cons~m~a~l))$ and $\tau l$
have type $(Vector~p)$. So,
for all $\eta$, $\eta \tau  (Tail~n~(Cons~m~a~l))$ and $\eta \tau l$ have
the same type and the rule is well-typed.

Therefore, sufficient condition for the rule $l \lra r$ to be
well-typed is that $l$ has a most general typing substitution $\tau$
and $\tau l$ and $\tau r$ have the same type. This leads to the rule
$$\irule{l\text{ is a pattern}~~~\Gamma,\Delta \vdash \tau l : T
         ~~~\Gamma,\Delta \vdash \tau r : T~~~dom(\Delta)\subseteq FV(l)}
        {\Gamma \vdash l \lra^\Delta r}
        {}$$
where $\tau$ is the most general typing substitution for $l$ in
$\Gamma, \Delta$.

This rule subsumes the first one as, if $l$ is well-typed, it has the
identity as a most general typing substitution.

\begin{lemma}[Theorem 3.5.8 in \cite{Saillard15}]
  \label{lem:wfwt}
  Let $\Gamma$ be a well-formed context satisfying the product
  compatibility.  If $\Gamma \vdash l \lra^\Delta r$, then
  $(l\lra^\Delta r)$ is well-typed in $\Gamma$.
\end{lemma}

Finally, we can prove the effectiveness theorem.

\begin{theorem}[Effectiveness, Theorem 4.5.10 and Theorem 6.3.1 in \cite{Saillard15}]
  If $\Gamma$ is a strongly well-formed context such that higher-order
  rewriting is confluent then $\Gamma$ satisfies the subject reduction
  property.  Moreover, if higher-order rewriting is terminating on
  well-typed terms, then typing is decidable.
\end{theorem}

\section{\textsc{Dedukti}}
\label{sec:dk}

\textsc{Dedukti} ({\tt http://dedukti.gforge.inria.fr/})
is a proof-checker for the $\lambda \Pi$-calculus
modulo theory. This system is not designed to develop proofs, but to
check proofs developed in other systems. In particular, it enjoys a
minimalistic syntax. It gives a concrete syntax to the $\lambda
\Pi$-calculus modulo theory, defined in Section \ref{sec:lpm}.

The typing algorithm relies on the confluence and termination of the
higher-order rewrite system together with $\beta$-reduction, to decide
whether two terms are convertible.

The text with a grey background is
executable \textsc{Dedukti} code. To avoid repetitions part of this
code is omitted, the complete version can be found on the
\textsc{Dedukti} website.

\subsection{A proof-checker for the  $\lambda \Pi$-calculus modulo
theory}

As in the $\lambda \Pi$-calculus modulo theory, there are two kinds of
constructs available in \textsc{Dedukti}:
variable declaration and rewrite rule declaration.  The only
available constant at the root of a file is \lst{Type}, that
corresponds to $Type$ above. The sort $Kind$ is not part of the input
syntax and is used only internally.

One only needs to declare the variables and rewrite rules, that will take
place in the \emph{global context}.
We can declare a variable \lst{nat}, together with
its type as follows
\begin{lstlisting}
nat : Type.
\end{lstlisting}

The $\Pi$-types $\Pi x: A~B$ are represented with a single-line arrow
\lst{x:A -> B}, which is thus a binder. When there is no dependency,
\lst{x} can be dropped, as in
\begin{lstlisting}
0 : nat.
S : nat -> nat.
\end{lstlisting}

Rewrite rules, together with their local context $\Delta$ are
declared in the form
\linebreak
\lst{[local context] l --> r.}
The types of the variables in the local context are inferred automatically. For instance
\begin{lstlisting}
def plus : nat -> nat -> nat.
[n] plus 0 n --> n.
[n1, n2] plus (S n1) n2 --> S (plus n1 n2).
\end{lstlisting}

A head symbol may be defined through several rewrite rules, in
this case, we allow their introduction all at once in the context,
instead of one by one, as is done in Section
\ref{subsec:lpm} and just above.
We express that the rules are defined all at once by writing just one
dot at the end of the sequence.
\begin{lstlisting}
[n] plus 0 n --> n
[n1, n2] plus (S n1) n2 --> S (plus n1 n2).
\end{lstlisting}

This has a direct impact on how the rules are typed. In the former
case, the first rule can be used to type the second, while
in the latter, it cannot.

Rewrite rules having the same head symbol need
not be added sequentially, they can be separated by an arbitrary
number of steps. In complex cases, the well-typing of a rewrite rule
on a symbol \lst{g} may depend on a rewrite rule on a symbol
\lst{f},
while further rewrite rules on \lst{f} may depend on those previously
declared on \lst{g}.

Moreover, one can add as many rules as wanted for a given head symbol, as
long as confluence is preserved. Checking confluence is out
of the scope of \textsc{Dedukti} itself, and is a separate concern. To
support confluence checking by external tools, an export feature
towards the old TPDB format~\cite{TPDB}, used in confluence competitions,
is available.
For instance, we can add the following rule to make \lst{plus} associative.
\begin{lstlisting}
[n1, n2, n3] plus n1 (plus n2 n3) --> plus (plus n1 n2) n3.
\end{lstlisting}

Defining a constant, standing for a complex expression, is identical
to defining a rewrite rule in an empty rewrite context. For
convenience, a special syntax is allowed
\begin{lstlisting}
def two := S (S 0).
\end{lstlisting}

Finally, $\lambda$-abstraction is expressed with the double arrow
\lst{=>}

\begin{lstlisting}
def K2 := x:nat => two.
\end{lstlisting}

Notice, that type annotation on the variable is optional
\cite{GBarMHei93}, when it can be reconstructed by bidirectional
typechecking
\cite{MBoeQCarOHer12,AChlLPetRHar05,TCoq96}. In this example, it is
mandatory.

The semantic difference between variables, constants and constructors
is purely in the eyes of the user, \textsc{Dedukti} treats them in the
same way.

It is also possible to split definitions in several files, that can be
imported. There also exist meta-directives to the checker, for
instance to ask the reduction of a term in head normal form.
Refer to the user manual \cite{usermanual}, for the details.

\subsection{Static and definable symbols}

We have seen in Section \ref{subsec:typingrewriterules}, that a key step
for accepting the linear rule
\begin{center}
\lst{tail n (Cons m a l) --> l}
\end{center}
is to compute the most general typing substitution for its
left-hand side, \lst{tail n (Cons m a l)}. The existence
of this substitution depends on the injectivity of the symbols
\lst{vector} and \lst{S}. So, \textsc{Dedukti} needs to be aware
that these symbols are injective with respect to the congruence $\eqbg$.
A simple way to ensure this property is to forbid rewrite rules
whose head symbol is \lst{vector} or \lst{S}.  Such a symbol,
whose appearance at
the head of rewrite rules is forbidden, is called a \emph{static} symbol
and is introduced by a declaration containing only the symbol name and its
type, such as \lst{vector} in the declarations
\begin{lstlisting}
A: Type.

vector: nat -> Type.
Nil : vector 0.
Cons : n : nat -> A -> vector n -> vector (S n).
\end{lstlisting}
Symbols such as \lst{tail} which may appear at the head of
rewrite rules are called \emph{definable} and are declared using the keyword
\lst{def}
\begin{lstlisting}
def tail: n:nat -> vector (S n) -> vector n.
\end{lstlisting}
Then, the rule
\begin{lstlisting}
[n, m, a, l] tail n (Cons m a l) --> l.
\end{lstlisting}
is accepted.

\subsection{Wildcards}

Instead of introducing new names for variables which will be inferred
by the unification step, the wildcard symbol \_ can be used in the
left-hand side as an unnamed pattern, matching any term, so we might write
\begin{lstlisting}
[n, a, l] tail n (Cons _ a l) --> l.
\end{lstlisting}
or even
\begin{lstlisting}
[l] tail _ (Cons _ _ l) --> l.
\end{lstlisting}

\subsection{Guards}

For the rare cases where the injectivity of a definable symbol would be
needed to solve a unification problem, or when it would be
preferable to use a non-normal term in the left-hand side of a rewrite
rule, \textsc{Dedukti} implements a {\em guard} mechanism.

For this example, the user can add brackets---the guard---around the
subterm whose shape is known by typing constraints.
\begin{lstlisting}
[n, a, l] tail n (Cons {n} a l) --> l.
\end{lstlisting}
\textsc{Dedukti} will type-check the rule as if there were no
brackets, but the rewrite rule declared will be the linear one, that
is to say the rule obtained by replacing the guard by a fresh
variable.

\textsc{Dedukti} cannot check in general the validity of the guard,
namely that it indeed follows from typing constraints, since it is
undecidable. Thus, in order to guarantee in any case a type-safe
reduction, \textsc{Dedukti} checks, each time the rule is used, that
the typing constraints are satisfied.  If they are not, this means that
the rewrite rule is not well-typed and \textsc{Dedukti} fails with an
error message.

\section{Constructive predicate logic and Deduction modulo theory}
\label{sec:constructive}

The first example of theory that can be embedded in \textsc{Dedukti}
is constructive predicate logic. As the type system is derived from
the $\lambda \Pi$-calculus, we shall see that those embeddings can be
defined quite naturally, and that, with rewriting, we can endow them
with a computational content and make them shallow.

\subsection{Minimal predicate logic}
\label{minimal}

Even without rewrite rules, the $\lambda \Pi$-calculus can express
proofs of minimal many-sorted predicate logic, that is the fragment of
constructive many-sorted predicate logic where the only connective is
the implication $\Rightarrow$ and the only quantifiers are the
universal quantifiers $\fa_s$, one for each sort.

\begin{definition}[Language embedding]
\label{deflangembedding1}
To each language ${\cal L}$ of predicate logic, we associate a context
$\Sigma$ containing
\begin{itemize}
\item for each sort $s$ of ${\cal L}$, a variable  \lst{s}, of type
\lst{Type},
\item for each function symbol $f$ of arity $\langle s_1, \ldots, s_n,
s'\rangle$, a variable \lst{f}, of type \linebreak
\lst{s}$_{\tt 1}$ \lst{-> ... -> s}$_{\tt n}$ \lst{-> s'},
\item for each predicate symbol $P$ of arity $\langle s_1, \ldots, s_n
\rangle$, a variable \lst{P}, of type \linebreak
\lst{s}$_{\tt 1}$ \lst{-> ... -> s}$_{\tt n}$ \lst{-> Type}.
\end{itemize}
\end{definition}

For instance
\begin{lstlisting}
s : Type.
P : s -> Type.
\end{lstlisting}

\begin{definition}[Term embedding]
\label{def:termembedding}
Let $t$ be a term in the language ${\cal L}$,
we express the term $t$ as a $\lambda$-term as
\begin{itemize}
\item $|x| = \mbox{\lst{x}}$,
\item $|f(t_1, \ldots, t_n)| = (\mbox{\lst{f}}~|t_1|~\ldots~|t_n|)$.
\end{itemize}
\end{definition}

The following lemma is proved with a simple induction on $t$.

\begin{lemma}
If $t$ is a term of sort $s$ in the language ${\cal L}$ and
$\Gamma$ a context containing for each variable $x$ of sort
$s'$ free in $t$, a variable \lst{x}, of type \lst{s'}, then
$\Sigma, \Gamma \vdash |t|:\mbox{\lst{s}}$.
\end{lemma}

\begin{definition}[Proposition embedding]
Let $A$ be a proposition in the language ${\cal L}$,
we express the proposition $A$ as the following $\lambda$-term
\begin{itemize}
\item $\|P(t_1,\ldots,t_n)\| = (\mbox{\lst{P}}~|t_1|~\ldots~|t_n|)$,
\item $\|A \Rightarrow B\| = \|A\|$ \lst{->} $ \|B\|$,
\item $\|\fa_s x~A\| =$ \lst{x:s ->} $\|A\|$.
\end{itemize}
\end{definition}

The following lemma is proved with a simple induction on $A$.
\begin{lemma}
If $A$ is a proposition in the language ${\cal L}$ and
$\Gamma$ a context containing for each variable $x$ of sort
$s$ free in $A$, a variable \lst{x}, of type \lst{s}, then
$\Sigma, \Gamma \vdash \|A\|:\mbox{\lst{Type}}$.
\end{lemma}

The following theorem is proved with a simple induction on proof structure
and on the structure of the normal form of $\pi$.
\begin{theorem}[Proof embedding]
Let ${\cal L}$ be a language and $A_1, \ldots, A_n \vdash B$ be a sequent
in ${\cal L}$, $\Gamma_1$ a context containing
for each variable $x$ of sort $s$ free in $A_1, \ldots, A_n \vdash B$,
a variable \lst{x}, of type \lst{s},
and $\Gamma_2$ be a context containing, for each hypothesis $A_i$,
a variable \lst{a}$_{\tt i}$, of type $\|A_i\|$. Then, the sequent
$A_1, \ldots, A_n \vdash B$ has a proof in Natural deduction, if and only if
there exists a $\lambda$-term $\pi$ such that $\Sigma, \Gamma_1, \Gamma_2
\vdash \pi:\|B\|$, if and only if
there exists a normal $\lambda$-term $\pi$ such that $\Sigma, \Gamma_1, \Gamma_2
\vdash \pi:\|B\|$.
\end{theorem}

For instance, the proposition $\fa_T x~(P(x) \Rightarrow P(x))$
is embedded as \lst{x:T -> (P x) -> (P x)} and
it has the proof \lst{x:T => a:(P x) => a}.

\begin{lstlisting}
def Thm : x:s -> (P x) -> (P x) := x:s => a:(P x) => a.
\end{lstlisting}

Note that if $B$ is a proposition, then the term $\|B\|$ is a type, in
the sense that it has type \lst{Type} and that it is the type of the
proofs of $B$. But not all
terms of type \lst{Type} have the form $\|B\|$, for instance if $s$ is
a sort of the language, then \lst{s} has type \lst{Type} but there exists
no proposition $B$ such that $\mbox{\lst{s}} = \|B\|$.  So the
concepts of proposition and type are not identified: propositions are
types, but not all types are propositions.

\subsection{Constructive predicate logic}
\label{constructive}

The $\lambda \Pi$-calculus has been designed to express functional
types but not Cartesian product types, disjoint union types, unit
types, or empty types. Thus, the connectives and quantifiers $\top$,
$\bot$, $\neg$, $\wedge$, $\vee$, and $\ex$ cannot be directly
expressed as $\Rightarrow$ and $\fa$ in Section~\ref{minimal}. One
possibility is to extend the $\lambda
\Pi$-calculus with Cartesian product types or inductive types. Another
is to use the rewrite rules of the $\lambda \Pi$-calculus modulo
theory to define these connectives and quantifiers.

We first modify the expression of minimal predicate logic presented in
Section~\ref{minimal}. We introduce a variable \lst{o}, of type
\lst{Type}
\begin{lstlisting}
o : Type.
\end{lstlisting}
and instead of expressing a predicate symbol as a term of type
\lst{s}$_{\tt 1} $\lst{-> ... -> s}$_{\tt n}$ \lst{-> Type},
we now express it as a term of type
\lst{s}$_{\tt 1}$ \lst{-> ... -> s}$_{\tt n}$ \lst{-> o}

\begin{lstlisting}
s : Type.
P : s -> o.
\end{lstlisting}
Thus, the atomic proposition $P(c)$ is expressed, in a first
step, as the term \lst{(P c)}, of type \lst{o}. We can also introduce a
variable \lst{imp} of type \lst{o -> o -> o}
\begin{lstlisting}
imp : o -> o -> o.
\end{lstlisting}
in order to express the
proposition $P(c) \Rightarrow P(c)$ as the term \lst{(imp (P c) (P c))},
of type \lst{o}.
Then, to make a connection to Section~\ref{minimal}, we introduce a variable
\lst{eps}---read ``epsilon''---of type \lst{o -> Type}
\begin{lstlisting}
def eps : o -> Type.
\end{lstlisting}
and the proposition $P(c)$ can be
expressed as the term \lst{eps (P c)}, of type \lst{Type}. The type
$o$ is thus a universe ``{\em à la} Tarski'' \cite{MartinLof}.
The
proposition $P(c) \Rightarrow P(c)$ can then be expressed both as
\lst{eps (imp (P c) (P c))}
and as
\lst{eps (P c)) -> (eps (P c)} and we add
the rewrite rule
\begin{lstlisting}
[x, y] eps (imp x y) --> (eps x) -> (eps y).
\end{lstlisting}
to identify these expressions.

This allows also to add other connectives and quantifiers, for instance the
conjunction \lst{and} as a variable of type \lst{o -> o -> o}, together
with the rewrite rule that express its meaning through the so called
``second-order'' expression of the conjunction
\begin{lstlisting}
and: o -> o -> o.
[x, y] eps (and x y) --> z:o -> (eps x -> eps y -> eps z) -> eps z.
\end{lstlisting}

\begin{definition}[Language embedding]
\label{deflangembedding2}
To each language ${\cal L}$ of predicate logic, we associate a context
$\Sigma$ containing
\begin{itemize}
\item for each sort $s$ of ${\cal L}$ a variable \lst{s} of type \lst{Type},
\item a variable \lst{o} of type \lst{Type},
\item for each function symbol $f$ of arity $\langle s_1, \ldots, s_n,
s'\rangle$, a variable \lst{f}, of type
\lst{s}$_{\tt 1}$\lst{ -> ... -> s}$_{\tt n}$\lst{ -> s'},
\item for each predicate symbol $P$ of arity $\langle s_1, \ldots, s_n
\rangle$, a variable \lst{P}, of type
\lst{s}$_{\tt 1}$\lst{ -> ... -> s}$_{\tt n}$\lst{ -> o},
\item variables \lst{top} and \lst{bot} of type \lst{o},
\item a variable \lst{not} of type \lst{o -> o},
\item variables \lst{imp}, \lst{and}, and \lst{or} of type \lst{o -> o -> o},
\item for each sort $s$ of ${\cal L}$ a variable \lst{fa_s} and a
 variable \lst{ex_s} of type \lst{(s -> o) -> o},
\item a variable \lst{eps} of type \lst{o -> Type}.
\end{itemize}
\end{definition}

The embedding of terms is defined as in Definition~\ref{def:termembedding}
and that of propositions comes in two steps. First, the translation
$|.|$ associates a term of type \lst{o} to each proposition. Then the symbol
\lst{eps} is applied to this term to obtain a term of type \lst{Type}.

\begin{definition}[Proposition embedding]
Let $A$ be a proposition in the language ${\cal L}$.
We express the proposition $A$ as a $\lambda$-term as
\begin{itemize}
\item $|P(t_1,\ldots,t_n)| = (\mbox{\lst{P}}~|t_1|~\ldots~|t_n|)$,
\item $|\top| = \mbox{\lst{top}}$, $|\bot| = \mbox{\lst{bot}}$,
\item $|\neg A| = \mbox{\lst{not}}~|A|$,
\item $|A \Rightarrow B| = \mbox{\lst{imp}}~|A|~|B|$,
$|A \wedge B| = \mbox{\lst{and}}~|A|~|B|$,
$|A \vee B| = \mbox{\lst{or}}~|A|~|B|$,
\item
$|\fa_s x~A| = $ \lst{fa_s (x:s =>} $|A|$\lst{)},
$|\ex_s x~A| = $ \lst{ex_s (x:s =>} $|A|$\lst{)},
\end{itemize}
and finally
\begin{itemize}
\item $\|A\| = $ \lst{eps} $|A|$.
\end{itemize}
\end{definition}

The following lemma is proved with a simple induction on $A$.
\begin{lemma}
If $A$ is a proposition in the language ${\cal L}$ and
$\Gamma$ a context containing for each variable $x$ of sort
$s$ free in $A$, a variable \lst{x}, of type \lst{s}, then
$\Sigma, \Gamma \vdash |A|:\mbox{\lst{o}}$
and
$\Sigma, \Gamma \vdash \|A\|:\mbox{\lst{Type}}$.
\end{lemma}

Finally, we add the rewrite rules expressing the meaning of the
connectives---\lst{and} is presented above---and quantifiers. The
rationale behind those rules is, that they turn the deep
\lst{o}-level encoding into a shallow \lst{Type}-level encoding.

\begin{lstlisting}
def top : o.
[ ] eps top --> z:o -> (eps z) -> (eps z).

def bot : o.
[ ] eps bot --> z:o -> (eps z).

def not : o -> o := x:o => imp x bot.

def or : o -> o -> o.
[x, y] eps (or x y)
        --> z:o -> (eps x -> eps z) -> (eps y -> eps z) -> eps z.

def fa_s : (s -> o) -> o.
[y] eps (fa_s y) --> x:s -> eps (y x).

def ex_s : (s -> o) -> o.
[y] eps (ex_s y) --> z:o -> (x:s -> eps (y x) -> (eps z)) -> eps z.
\end{lstlisting}



\begin{theorem}[Embedding of proofs, \cite{Dorra}]
\label{theoremconstructive}
Let ${\cal L}$ be a language, $\Sigma$ be
the associated context, $V$ be a set of variables and $A_1$, \ldots,
$A_n$, $B$ be propositions in ${\cal L}$ such that the free variables
of $A_1, \ldots, A_n$, $B$ are in $V$.  Let $\Gamma_1$ be a context
containing for each variable $x$ of sort $s$ in $V$, a variable
\lst{x}, of type \lst{s}, and $\Gamma_2$ be a context containing, for
each hypothesis $A_i$ a variable \lst{a}$_{\tt i}$ of type $\|A_i\|$.

Then, the sequent
$A_1, \ldots, A_n \vdash B$ has a proof in Natural deduction if and only if
there exists a $\lambda$-term $\pi$ such that
$\Sigma, \Gamma_1, \Gamma_2 \vdash \pi:\|B\|$.
\end{theorem}

Again, not all terms of type \lst{Type} have the form $\|B\|$, only those
convertible to a term of the form \lst{(eps p)}, where \lst{p} is a term
of type \lst{o}, do. For instance \lst{o} has type \lst{Type}, but
there is no proposition $B$ such that \lst{o} $= \|B\|$.
In contrast, every normal term of type \lst{o} is equal to $|B|$ for some
$B$.  Thus, the type \lst{o}, also called \lst{Prop} or \lst{bool} in
many systems, is the type of propositions.

So, the concepts of proposition and type are not identified in this
encoding: the type \lst{o} of propositions is a subtype of the type of
all types, \lst{Type}, through the explicit embedding \lst{eps}, of
type \lst{o -> Type}. This way \lst{o} can have type \lst{Type},
without introducing inconsistencies.

\subsection{Deduction modulo theory}
\label{deduction-modulo}

The embedding presented in Section \ref{constructive} extends readily
to Deduction modulo theory: a theory is just expressed by declaring
its rewrite rules.

Moreover, some theories, such as arithmetic or
set theory, contain a sort $\iota$ and a sort $\kappa$ for the classes
of elements of sort $\iota$ and a comprehension scheme
$$\fa x_1~\ldots~\fa x_n \ex c~\fa y~(y \in c \Leftrightarrow A)$$
or its skolemized form
$$\fa x_1~\ldots~\fa x_n~\fa y~
(y \in f_{x_1, \ldots, x_n, A}(x_1, \ldots, x_n) \Leftrightarrow A)$$
In the $\lambda \Pi$-calculus modulo theory this sort can be omitted and
replaced by the type \lst{i -> o}.

For instance, Heyting arithmetic \cite{DWarith,Allali} can be
presented in Deduction modulo theory with nine rewrite rules defining
addition, multiplication, equality, and numberness
\begin{lstlisting}
nat : Type.

0 : nat.
S : nat -> nat.
def plus : nat -> nat -> nat.
def times : nat -> nat -> nat.
def equal : nat -> nat -> o.
N : nat -> o.
[y] plus 0 y --> y
[x, y] plus (S x) y --> S (plus x y).
[y] times 0 y --> 0
[x, y] times (S x) y --> plus (times x y) y.
[ ] equal 0 0 --> top
[x] equal (S x) 0 --> bot
[y] equal 0 (S y) --> bot
[x, y] equal (S x) (S y) --> equal x y.
[n] eps (N n) -->
  k:(nat->o) -> eps (k 0) ->
  eps ( fa_nat (y:nat => imp (N y) (imp (k y) (k (S y))) ) ) ->
  eps (k n).
\end{lstlisting}

The last rewrite rule, accounting for numberness, expresses that any
\lst{n}, of type \lst{nat}, is a natural number if it verifies
all predicates \lst{k}, verified by $0$ and closed by successor.
Then we can prove the proposition $\fa x~N(x)
\Rightarrow (x + 0) = x$
\begin{lstlisting}
def tt : eps top := z:o => p:eps z => p.

def k := x:nat => equal (plus x 0) x.

def z_r_neutral : eps ( fa_nat ( x => imp (N x) (k x)))
                := x:nat => p:eps (N x) =>
                   p k tt (y:nat => q: eps (N y) => r:eps (k y) => r).
\end{lstlisting}

\subsection{\textsc{iProverModulo}}

In this section we present the \textsc{Dedukti} back-end of
\textsc{iProverModulo}, more information can be found about
\textsc{iProverModulo} in~\cite{burel11experimenting} and about the
back-end in~\cite{Burel13}.

\textsc{iProver} \cite{korovin08iprover} is a proof-search
system for classical predicate logic, based on ordered resolution with
selection \cite{bachmair01resolution}.  An optional
Instantiation-Generation rule \cite{ganzinger03new} can also be used,
but we shall use a version of \textsc{iProver} where this option
has not been activated.

As usual, a literal is either an atomic proposition or the negation of
an atomic proposition and a clause is a set of literals.  If $C =
\{L_1,\dots,L_n\}$, we write $\ulcorner C \urcorner$ for the
proposition $\forall x_1~\dots~\forall x_m~(L_1\vee (\dots
\vee(L_{n-1} \vee L_n)\dots))$ where $x_1, \dots, x_m$ are the
variables of $C$.  To prove a proposition $A$, its negation $\neg A$ is
first transformed into a set of clauses $\{C_1, ..., C_n\}$ by an external
tool. The proof then amounts to deriving the
empty clause, using the usual resolution and
factoring rules. The only difference between standard Resolution and
ordered resolution with selection is that, in this second method,
these rules are restricted to reduce the search space, while preserving
completeness.

\textsc{iProver} has been extended to Deduction modulo theory, leading
to \textsc{iProverModulo}~\cite{burel11experimenting}.  Proposition
rewrite rules are simulated using so-called one-way clauses
\cite{polar} that are handled with just more restrictions on the
Resolution rule.  Term rewrite rules, in contrast, lead to introduce
an additional Narrowing rule.

When the proposition $\neg A$ is transformed into a set of clauses
$C_1, ..., C_n$ and the empty clause can be derived from $C_1, ...,
C_n$ with the Resolution, Factoring, and Narrowing rules, a proof in
Deduction modulo theory of the sequent $\ulcorner C_1 \urcorner, ...,
\ulcorner C_n \urcorner \vdash \bot$ can be built, and this proof can
then be transformed into a proof of the sequent $\vdash A$. The latter
transformation is only possible in classical logic, but the proof of
the sequent $\ulcorner C_1 \urcorner, ..., \ulcorner C_n \urcorner
\vdash \bot$ itself is constructive. And indeed, since version
0.7-0.2, \textsc{iProverModulo} has an export functionality that
builds a \textsc{Dedukti} proof of $\ulcorner C_1 \urcorner, ...,
\ulcorner C_n \urcorner \vdash \bot$, every time it derives the empty
clause from $C_1, ..., C_n$.

A clause $C$ can be transformed into a proposition $\ulcorner C
\urcorner$ and then to a \textsc{Dedukti} term $\| \ulcorner C
\urcorner \|$, of type \lst{Type}, as seen in Section
\ref{constructive}.  However, instead of using the binary
disjunction~$\vee$, we generalize it to a multiary disjunction and
translate directly the clause $C = \{L_1, ..., L_n\}$ as the
\textsc{Dedukti} term $\llbracket C \rrbracket$ of, type \lst{Type},
as
$$
\mbox{$\llbracket C \rrbracket =$
\lst{x}$_{\tt 1}$\lst{:i->...->x}$_{\tt m}$\lst{:i->(}$\|L_1\|$\lst{->(eps bot))->...->(}$\|L_n\|$\lst{ ->(eps bot))->(eps bot)}}
$$
where we use the sort \lst{i} for terms. In particular, the
translation of the empty clause is $\llbracket \varnothing \rrbracket
= \mbox{\lst{eps bot}} = \|\bot\|$.

Proofs built by
\textsc{iProverModulo} are then expressed in \textsc{Dedukti}
step by step. We first declare a variable \lst{d}$_{\tt C}$, of type
$\|\ulcorner C \urcorner\|$ for each clause $C$ in the set $C_1, ..., C_n$.
\begin{center}
\lst{d}$_{\tt C}$\lst{:eps}~$|\forall x_1~\dots\forall x_m~(L_1\vee \dots\vee(L_{n-1}\vee L_n)\dots)|$
\end{center}
Then, using these variables, we can define a term \lst{c}$_{\tt C}$,
of type $\llbracket C \rrbracket$.
For each new clause generated by the inference rules, we
build a proof of the conclusion from the proof
of the premises. For instance, for the Resolution rule
$$\irule{C \cup \{P\}
         ~~~~~~~~~
         D\cup\{\neg Q\}}
  {\sigma(C\cup D)}
  {}$$
where $\sigma$ is the most general unifier of $P$ and $Q$,
we have a term \lst{c} of type $\llbracket C\cup \{P\}\rrbracket$,
and a term \lst{d} of type $\llbracket D\cup \{\neg Q\}\rrbracket$.
 We can combine them to define a term of the resolvant type
 $\llbracket \sigma(C\cup D)\rrbracket$.
See \cite{Burel13} for more details. As an example, the proof

$$\hspace*{-5cm}
\irule{(1)
         ~~~~~~~~~~~~~~~~~~~~~~~~~~~~~~~~~~~~~~~~~~~~
         \irule{\irule{\{ P(a,x), P(y,b), P(y,x) \}}
                      {P(a,b)~~(1)}
                      {\mbox{Factoring}}
                ~~~~~~~~~~~~~~~~~~~~~~~~~~
                \{ \neg P(a,x), \neg P(y,b) \}
         }
               {\neg P(a,x)}
               {\mbox{Resolution}}
        }
        {\varnothing}
        {\mbox{Resolution}}$$
is a derivation of the empty clause from the clauses $C_1 = \{ P(a,x),
P(y,b), P(y,x)\}$ and
\linebreak
$C_2 = \{ \neg P(a,x), \neg P(y,b) \}$. It is
translated in \textsc{Dedukti} as
\begin{lstlisting}
i : Type.

P: i -> i -> o.
a : i.
b : i.
d1: eps (fa_i (x => fa_i (y => or (P a x) (or (P y b) (P y x))))).
d2: eps (fa_i (x => fa_i (y => or (not (P a x)) (not (P y b))))).
def c1 : x : i -> y : i -> (eps (P a x) -> eps bot) ->
                           (eps (P y b) -> eps bot) ->
                           (eps (P y x) -> eps bot) -> eps bot
       := x => y => l1 => l2 => l3 => z =>
          d1 x y z (l1' => l1 l1' z)
             (sb1 => sb1 z (l2' => l2 l2' z) (l3' => l3 l3' z)).
def c2 : x : i -> y : i ->
          ((eps (P a x) -> eps bot) -> eps bot) ->
          ((eps (P y b) -> eps bot) -> eps bot) -> eps bot
       := x => y => l1 => l2 => z =>
          d2 x y z (l1' => l1 l1' z) (l2' => l2 l2' z).
def c3 : (eps (P a b) -> eps bot) -> eps bot
       := l1 => c1 b a l1 l1 l1.
def c4 : x : i -> ((eps (P a x) -> eps bot) -> eps bot) -> eps bot
       := x => l1 => c3 (tp => c2 x a l1 (tnp => tnp tp)).
def c5 : eps bot := c3 (tp => c4 b (tnp => tnp tp)).
\end{lstlisting}
where \lstinline!d1! (resp.\ \lstinline!d2!) is the declaration of $\| \ulcorner C_1
\urcorner \|$ (resp.\ $\| \ulcorner C_2
\urcorner \|$); \lstinline!c1! (resp.\ \lstinline!c2!) its translation
using a multiary disjunction $\llbracket C_1 \rrbracket$ (resp.\ $\llbracket C_2 \rrbracket$); and \lstinline!c3!, \lstinline!c4! and
\lstinline!c5! correspond to the clauses derived from $C_1$ and $C_2$ in
the proof tree above.






\textsc{iProverModulo} can be used to produce \textsc{Dedukti} proofs
for 3383 problems of the TPTP problem library v6.3.0~\cite{SS98}. The
generated \textsc{Dedukti} files weight a total of 38.1 MB, when
gzipped.

\section{Classical predicate logic}
\label{sec:classical}

A logical framework should be able to express both constructive and
classical theories. A way to mix constructive and classical arguments
is to apply the adjectives ``constructive'' and ``classical'' not to
proofs, but to connectives. So, we shall not say that the proposition
$P \vee \neg P$ has a classical proof and no constructive one. Instead,
we shall say that the proposition $P \vee_{\mathrm c} \neg_{\mathrm c} P$ has a proof and
the proposition $P \vee \neg P$ does not, distinguishing the classical
connectives $\vee_{\mathrm c}$, $\neg_{\mathrm c}$, etc. from the constructive ones
$\vee$, $\neg$, etc.

Reaping the benefit of the work done on negative translations by
Kolmogorov, G\"odel, Gentzen, Kuroda, and others, we want to define
the classical connectives from the constructive ones, adding double
negations before or after the connectives. For instance, defining $A
\vee_{\mathrm c} B$ as $\neg \neg (A \vee B)$.  This approach has been
investigated by Hermant, Allali, Dowek, Prawitz, etc.
\cite{HermantAllali,Doweknotnot,Prawitz}. In all these
works the main problem is that the translation of an atomic
proposition $P$ needs to be $\neg \neg P$ and in this case there is no
connective to which these negations can be incorporated. This has lead to
various solutions: considering two different provability predicates
\cite{HermantAllali}, considering two entailment relations
\cite{Doweknotnot}, incorporating the double negation to the predicate
symbol at the root of $P$ \cite{Prawitz}.

We consider here another solution that fits better our goal to express
classical proofs in a logical framework: the introduction of a new
connective.

\subsection{Classical connectives and quantifiers}

The usual syntax for predicate logic distinguishes terms and propositions
$$
\begin{array}{c}
  \begin{array}{r@{~~~::=~~~}l}
    t & x~|~f(t,\ldots,t)
  \end{array}\\
  A~~~=~~~P(t_1, \ldots, t_n)~|~\top~|~\bot~|~\neg A~|~A \Rightarrow
  A~|~A \wedge A~|~A \vee A~|~\fa x~A~|~\ex x~A
\end{array}
$$
The syntactic category of terms introduces only function symbols which are
specific to the theory, while the second syntactic category of propositions
introduces both predicate symbols, which are specific to the theory, and
connectives which are not.

A clearer presentation completely separates the symbols of the theory
and those of the logic introducing a third syntactic category for atoms
$$
\begin{array}{c}
  \begin{array}{r@{~~~::=~~~}l}
  t & x~|~f(t,\ldots,t)\\
  a & P(t_1, \ldots, t_n)
  \end{array}\\
  A~~=~~a~|~\top~|~\bot~|~\neg A~|~A \Rightarrow A~|~A \wedge A~|~A \vee A
  ~|~\fa x~A~|~\ex x~A
\end{array}
$$

When formalizing such a definition in a logical framework we must introduce
a type for terms of each sort, a type for atoms, and a type for propositions,
and we cannot just say that atoms are propositions.
Instead, we must introduce an explicit embedding from atoms to propositions,
a new connective, $\ato$, so the propositional category becomes
$$
A~~~::=~~~\ato a~|~\top~|~\bot~|~\neg A~|~A \Rightarrow A
~|~A \wedge A~|~A \vee A~|~\fa x~A~|~\ex x~A
$$

This way, the new connective $\ato$, in its classical version, may
incorporate the double negation usually put on the atom itself.
For instance, Kolmogorov translation yields the following definitions of
the classical atomic, conjunction, and disjunction connectives
$$
\begin{array}{rcl}
  \ato_{\mathrm c} A & = & \neg \neg \ato A\\
  A \land_{\mathrm c} B & = & \neg \neg (A \land B)\\
  A \lor_{\mathrm c} B & = & \neg \neg (A \lor B) \\
\end{array}
$$
Assuming the constructive connectives and quantifiers of
Section~\ref{constructive}, the full set of classical connectives and
quantifiers can now be defined in \textsc{Dedukti} as
\begin{lstlisting}
alpha : Type.
def atom : alpha -> o.
def atom_c : alpha -> o := p: alpha => not (not (atom p)).
def top_c : o := not (not top).
def bot_c : o := not (not bot).
def not_c : o -> o := x:o => (not (not (not x))).
def and_c : o -> o -> o := x:o => y:o => not (not (and x y)).
def or_c  : o -> o -> o := x:o => y:o => not (not (or x y)).
def imp_c : o -> o -> o := x:o => y:o => not (not (imp x y)).
def fa_s_c : (s -> o) -> o := x:(s -> o) => not (not (fa_s x)).
def ex_s_c : (s -> o) -> o := x:(s -> o) => not (not (ex_s x)).
\end{lstlisting}

The embedding of languages is as in Definitions
\ref{deflangembedding1} and \ref{deflangembedding2}, except that the
predicate symbol now have a target type \lst{alpha}, instead of
\lst{Type}, or \lst{o}.  The embedding of terms is defined as in
Definition~\ref{def:termembedding}.  We define the constructive and
classical embeddings for propositions as follows.

\begin{definition}[Proposition embedding]
Let $A$ be a proposition in the language ${\cal L}$
we embed it constructively as a $\lambda$-term as
\begin{itemize}
\item $|P(t_1,\ldots,t_n)| =$ \lst{atom
  (P}$~|t_1|~\ldots~|t_n|$\lst{)},
\item $|\top| = \mbox{\lst{top}}$, $|\bot| = \mbox{\lst{bot}}$,
\item $|\neg A| = \mbox{\lst{not}}~|A|$,
\item $|A \Rightarrow B| = \mbox{\lst{imp}}~|A|~|B|$,
$|A \wedge B| = \mbox{\lst{and}}~|A|~|B|$,
$|A \vee B| = \mbox{\lst{or}}~|A|~|B|$,
\item
$|\fa_s x~A| = $ \lst{fa_s (x:s =>} $|A|$\lst{)},
$|\ex_s x~A| = $ \lst{ex_s (x:s =>} $|A|$\lst{)}.
\end{itemize}
We embed $A$ classically as a $\lambda$-term as
\begin{itemize}
\item $|P(t_1,\ldots,t_n)|_{\mathrm c} =$ \lst{atom_c (P}
  $|t_1|~\ldots~|t_n|$\lst{)},
\item $|\top|_{\mathrm c} = \mbox{\lst{top_c}}$, $|\bot|_{\mathrm c} = \mbox{\lst{bot_c}}$,
\item $|\neg A|_{\mathrm c} = \mbox{\lst{not_c}}~|A|_{\mathrm c}$,
\item $|A \Rightarrow B|_{\mathrm c} = \mbox{\lst{imp_c}}~|A|_{\mathrm c}~|B|_{\mathrm c}$,
$|A \wedge B|_{\mathrm c} = \mbox{\lst{and_c}}~|A|_{\mathrm c}~|B|_{\mathrm c}$,
$|A \vee B|_{\mathrm c} = \mbox{\lst{or_c}}~|A|_{\mathrm c}~|B|_{\mathrm c}$,
\item
$|\fa_s x~A|_{\mathrm c} = $ \lst{fa_s_c (x:s =>} $|A|_{\mathrm c}$ \lst{)},
$|\ex_s x~A|_{\mathrm c} = $ \lst{ex_s_c (x:s =>} $|A|_{\mathrm c}$ \lst{)}.
\end{itemize}
\end{definition}

The following lemma is proved with a simple induction on $A$.

\begin{lemma}
If $A$ is a proposition in the language ${\cal L}$ and
$\Gamma$ a context containing for each variable $x$ of sort
$s$ free in $A$, a variable \lst{x}, of type \lst{s}, then
$\Sigma, \Gamma \vdash |A|:\mbox{\lst{o}}$
and
$\Sigma, \Gamma \vdash |A|_{\mathrm c}:\mbox{\lst{o}}$.
\end{lemma}

Note that the normal form of $|A|_{\mathrm c}$ is the result of the Kolmogorov
translation of $A$, constructively embedded with $|~|$. Thus,
combining Kolmogorov's results about his translation with Theorem
\ref{theoremconstructive}, we get the following theorem.

\begin{theorem} A proposition $A$ has a proof in constructive logic
  if and only if the type \lst{eps} $|A|$ is inhabited. A proposition
  $A$ has a proof in classical logic if and only if the type
  \lst{eps} $|A|_{\mathrm c}$ is inhabited.
\end{theorem}

For instance, the proposition
$$\ato_{\mathrm c} P \vee_{\mathrm c} \neg_{\mathrm c} \ato_{\mathrm c} P$$
has the following proof, where a redex has been introduced
\begin{lstlisting}
def lem : p : alpha -> eps (or_c (atom_c p) (not_c (atom_c p))) :=
 p => h0 : (eps (or (atom_c p) (not_c (atom_c p))) -> eps bot) =>
 (h1 : eps (not (atom_c p)) =>
  h0 (z =>
      h_left =>
      h_right =>
      h_right (h => h h1)))
 (h2 : eps (atom_c p) =>
  h0 (z =>
      h_left =>
      h_right =>
      h_left h2)).
\end{lstlisting}

Note that, since the symbols $\vee$ and $\vee_{\mathrm c}$, $\ato$ and $\ato_{\mathrm c}$, etc.
have different meanings the propositions
$$
\begin{array}{c}
  \fa x\fa y \fa z~(\ato~x \in \{y, z\} \Leftrightarrow (\ato~x = y \vee
  \ato~x = z))\\
  \fa_{\mathrm c} x\fa_{\mathrm c} y \fa_{\mathrm c} z~(\ato_{\mathrm c}~x \in \{y, z\} \Leftrightarrow_{\mathrm c}
  (\ato_{\mathrm c}~x = y \vee_{\mathrm c} \ato_{\mathrm c}~x = z))
\end{array}
$$
also have different meanings. In a mixed logic containing both
constructive and classical connectives, we can choose to take one, or
the other, or both as axioms, leading to different theories.

\subsection{Classical Deduction modulo theory}

Like constructive logic, classical logic can be extended with
a congruence defined by rewrite rules.
The translation of a rewrite rule of classical Deduction modulo theory
such as
$$\ato_{\mathrm c} x \in \{y, z\} \lra (\ato_{\mathrm c} x = y) \vee_{\mathrm c} (\ato_{\mathrm c} x = z)$$
could be expressed directly in \textsc{Dedukti}, but it would
introduce a critical pair because $\ato_{\mathrm c} P$ also reduces to $\neg
\neg \ato P$. To close this critical pair, we rather add the rule
$$\ato x \in \{y, z\} \lra (\ato_{\mathrm c} x = y) \vee (\ato_{\mathrm c} x = z)$$
that subsumes the first one
as $\neg \neg \ato  x \in \{y, z\}$ then rewrites to
$\neg \neg ((\ato_{\mathrm c} x = y) \vee (\ato_{\mathrm c} x = z))$.

Thus, to avoid critical pairs when
translating a rewrite rule of classical Deduction modulo theory, we must
remove the two head negations that would otherwise appear at the head
of the left-hand and right-hand sides of the rewrite rule. Said
otherwise, we turn the head connective of the left- and right-hand
sides into a constructive one.

We had two versions of the pairing axiom above, one constructive and
the other classical. In the same way,
this rule is different from the translation of the rewrite rule
of constructive Deduction modulo theory
$$\ato x \in \{y, z\} \lra (\ato x = y) \vee (\ato x = z)$$

\subsection{\textsc{Zenon}}
\label{zenon}

In this section we present \textsc{Zenon} 0.8.2 and
\textsc{Zenon Modulo} 0.4.2, more
information about \textsc{Zenon} can be found at
\cite{DBLP:conf/lpar/BonichonDD07}, and about \textsc{Zenon Modulo} at
\cite{Zenon-Modulo,bury:hal-01204701}.

\textsc{Zenon} is a proof search system for classical predicate
logic with equality, based on a tableaux method,
that is a sequent calculus with
one-sided sequents, where all propositions are on left-hand side.
\textsc{Zenon Modulo} is an
extension of \textsc{Zenon} that builds proofs in
Deduction modulo theory.  A proof of a
proposition $G$ in a context $\Gamma$ is a tableaux proof of the
sequent $\Gamma, \neg G \vdash$.

As we have seen, a proposition $G$ has a classical proof, if and only
if the type \lst{(eps} $|G|_{\mathrm c}$\lst{)} is inhabited in
\textsc{Dedukti}.  When proving a proposition $G$, \textsc{Zenon Modulo}
produces a
\textsc{Dedukti} proof-term, of type \lst{(eps} $|G|_{\mathrm c}$\lst{)}
\cite{CauderlierHalmagrand2015}.

More precisely it first produces a
\textsc{Dedukti} term of type
\lst{eps (not_c} $|G|_{\mathrm c}$\lst{) -> eps bot}.
This term can then be turned into a term of type
\lst{(eps} $|G|_{\mathrm c}$\lst{)} by applying the constructive theorem
$\neg \neg \neg A \Rightarrow \neg A$ twice, because $|G|_{\mathrm c}$ starts with a
negation.

These proofs of \lst{eps (not_c} $|G|_{\mathrm c}$\lst{) -> eps bot} are
generated by stating in \textsc{Dedukti} one lemma per tableaux rule.
For instance, the tableaux rule
$$\irule{A \vee_{\mathrm c} B}
        {A~~~~~~~~~B}
        {\mbox{Ror}}$$
that corresponds to the sequent calculus rule
$$\irule{\Gamma, A \vdash~~~\Gamma, B \vdash}
        {\Gamma, A \vee_{\mathrm c} B \vdash}
        {\mbox{$\vee$-left}}$$
is reflected by the following \textsc{Dedukti} lemma
\begin{lstlisting}
def Ror : A : o -> B : o ->
          (eps A -> eps bot) ->
          (eps B -> eps bot) ->
          eps (or_c A B) -> eps bot
   := A => B => HNA => HNB => HNNAB => HNNAB (HAB => HAB bot HNA HNB).
\end{lstlisting}

The translation of \textsc{Zenon Modulo} proofs to \textsc{Dedukti} has been
tested using a benchmark of 9,994 B Method~\cite{B-Book} proof
obligations, representing 5.5 GiB of input files.  The generated
\textsc{Dedukti} files weigh a total of 595 MB, when gzipped.


Around fifty axioms of the B Method set theory were turned into rewrite rules.
For instance, membership to the union between two sets is defined as
follows

\begin{lstlisting}
def mem : s -> s -> alpha.

def union : s -> s -> s.
[a, b, x] atom (mem x (union a b))
            --> (or (atom_c (mem x a)) (atom_c (mem x b))).
\end{lstlisting}

\section{Simple type theory}
\label{sec:simple-type-theory}

\subsection{Expressing simple type theory}
\label{simple-type-theory}

There are two ways to express Simple type theory \cite{Church40}, also
known as Higher-order logic, in the $\lambda \Pi$-calculus modulo
theory. First, Simple type theory can be
expressed in Deduction modulo theory \cite{DHKHOL}. Thus, as seen in
Section \ref{deduction-modulo}, it can be expressed in the $\lambda
\Pi$-calculus modulo theory. It can also be expressed as a Pure
type system \cite{Geuversthesis,Geuvers95}, hence in the $\lambda
\Pi$-calculus modulo theory \cite{CousineauDowek}, which leads to a
similar result.

In Simple type theory, the sorts of the logic are the types of the
simply-typed $\lambda$-calculus, with two basic types: $\iota$ for the
objects and $o$ for the propositions.  The language contains the
variables $\Rightarrow$, of type $o \to o \to o$, that represents
implication and $\forall_A$, for each simple type $A$, that represent
universal quantification. In order to avoid declaring an infinite
number of symbols, we represent simple types as \textsc{Dedukti} terms
of a type \lst{type} and pass elements of this type, to index a single symbol
$\fa$.

\begin{lstlisting}
type : Type.
o : type.
i : type.
arrow : type -> type -> type.

def term : type -> Type.
imp : term o -> term o -> term o.
forall : a : type -> (term a -> term o) -> term o.
\end{lstlisting}
One should not confuse \lst{type}, which is the type of objects
representing types of the Simple type theory, with \lst{Type}, which
is the type of \textsc{Dedukti} types. We endow the arrow types of the
\lst{type} level, the behavior of arrow types of the \lst{Type} level,
by adding the rule
\begin{lstlisting}
[a, b] term (arrow a b) --> term a -> term b.
\end{lstlisting}

\begin{definition}
The terms of Simple type theory are embedded as \textsc{Dedukti} terms
as follows
\begin{itemize}
\item $|x| =$ \lst{x},
\item $|\Rightarrow| =$ \lst{imp},
\item $|\forall_A| = $ \lst{forall} $|A|$,
\item $|M~N| = |M|~|N|$,
\item $|\lambda x:A~M| = $ \lst{x:} \lst{(term} $|A|$\lst{) =>} $|M|$.
\end{itemize}
\end{definition}

As in Section~\ref{constructive}, we declare a variable \lst{eps} that
maps a proposition to the type of its proofs, and we define its
meaning with the rewrite rules
\begin{lstlisting}
def eps : term o -> Type.

[p, q] eps (imp p q) --> eps p -> eps q.
[a, p] eps (forall a p) --> x : term a -> eps (p x).
\end{lstlisting}

This signature---call it $\Sigma_{\mathrm{STT}}$---defines the logic of minimal
constructive Simple type theory. For any proposition $A$, we define $\|A\| =
\mbox{\lst{eps}}~|A|$, and for any context $\Gamma = A_1,\ldots,A_n$, we
define $\|\Gamma\| =
h_1:\|A_1\|,\ldots,h_n:\|A_n\|$, where $h_1, ..., h_n$ are new variables.
Then we can express all the proofs of
Simple type theory.
\begin{theorem}[Theorem 5.2.11 and Theorem 6.2.27 in \cite{AssafThese}]
$\Gamma \vdash A$ is provable in Simple type theory, if and only if
  there is a $\lambda$-term $\pi$ such that
  $\Sigma_{\mathrm{STT}},\|\Gamma\| \vdash \pi : \|A\|$.
\end{theorem}
Moreover, the term $\pi$ is a straightforward encoding of the original
proof tree.

\subsection{{\textsc{HOL Light}} proofs}

The embedding of the previous Section~\ref{simple-type-theory} can be
adapted to represent variations of Simple type
theory, in particular classical ones like the system $\textrm{Q}_0$
\cite{Andrews86}, which is based on equality as a primitive connective
and which is used in modern implementations of Simple type theory,
like \textsc{HOL Light} \cite{AssafBurel2015}.

The system \textsc{HOLiDe}---pronounced ``holiday''---allows for the expression
\textsc{HOL Light} proofs in \textsc{Dedukti}, building on the Open
theory project \cite{Opentheory}.  \textsc{HOLiDe} supports all the
features of \textsc{HOL Light}, including prenex polymorphism, constant
definitions, and type definitions. It is able to express all of the
10MB OpenTheory standard theory library. This translation
results in a \textsc{Dedukti} library of 21.5 MB, when gzipped.
It could also be used to check in  \textsc{Dedukti} proofs developed
in other implementations of HOL, provided they can export proofs
to the Open theory format.

\section{Programming languages}
\label{sec:programming-languages}

Since rewriting is Turing-complete, we can also use the $\lambda
\Pi$-calculus modulo theory as a programming language and embed
programming languages into it. This allows to express proofs of programs
in \textsc{Dedukti}.  We target a shallow embedding in the logic, so
that we can reason about them as functions. In particular we cannot
speak about the number of lines of code of a program, or the number of
execution steps, but we can speak about functional properties of these
programs.

The embedding is defined by a rewrite system expressing the
operational semantics of the source language.

In this section, we present three examples, the $\lambda$-calculus,
the $\varsigma$-calculus and ML and then \textsc{FoCaLiZe} that is
a logic to reason about programs.

\subsection{Lambda-calculus}

For example, the untyped
$\lambda$-calculus can be encoded by
\begin{itemize}
\item $|x| = $ \lst{x},
\item $|t~u| =$ \lst{app} $|t|~|u|$,
\item $|\lambda x~t| = $ \lst{lam (x:term =>} $|t|$\lst{)}.
\end{itemize}
in the signature
\begin{lstlisting}
Term : Type.
lam : (Term -> Term) -> Term.
def app : Term -> Term -> Term.

[f,t] app (lam f) t --> f t.
\end{lstlisting}

Encoding programming languages in the $\lambda \Pi$-calculus modulo
theory and translating their libraries is
mandatory to check proofs of certified programs in \textsc{Dedukti}.
It is especially adequate for checking partial correctness
proofs---that implicitly assume termination of the program. As shallow
embeddings preserve the programming language binding, typing, and
reduction, we can abstract on the program reduction and only consider
programs as mathematical functions.

Obviously, preserving the programming language reduction can lead to a
non-terminating rewrite system, which is not effective in the
sense of Section~\ref{subsec:effective}. However it is not often a
problem in practice for the following reasons.
\begin{itemize}
\item The soundness of \textsc{Dedukti}, that is the fact that
\textsc{Dedukti} accepts well-typed terms only,
 does not depend on termination but only
  on product
\linebreak
compatibility---see Lemma \ref{lem:wfwt}.
\item The only place where the type-checking algorithm uses
  reduction is the conversion rule, to check convertibility of
  two types; as long as we do not use types depending on
  non-terminating terms, \textsc{Dedukti} terminates.
\item Even in presence of types depending on weakly-terminating terms,
  \textsc{Dedukti}'s strategy---comparison of weak head normal
  forms---often terminates.
\end{itemize}

For instance,
we can define a fixpoint operator in
simply-typed $\lambda$-calculus
\begin{lstlisting}
type : Type.
arrow : type -> type -> type.
def term : type -> Type.
[A,B] term (arrow A B) --> term A -> term B.

def fix : A : type -> B : type ->
          ((term A -> term B) -> term A -> term B) ->
          term A -> term B.
[A,B,F,a] fix A B F a --> F (fix A B F) a.
\end{lstlisting}
and prove properties of recursive terminating functions defined using
this \lst{fix} operator, such as the function \lst{mod2} defined below.

\begin{lstlisting}
nat : type.
def Nat := term nat.
O : Nat.
S : Nat -> Nat.
def match_nat : Nat -> Nat -> (Nat -> Nat) -> Nat.
[n0,nS]   match_nat O n0 nS --> n0
[n,n0,nS] match_nat (S n) n0 nS --> nS n.

def mod2 := fix nat nat
                (m2 => n =>
                  match_nat n
                    O
                    (p => match_nat p
                       (S O)
                       (q => m2 q))).

#CONV mod2 (S (S (S (S O)))), O.
\end{lstlisting}

The \lstinline{#CONV} command in last line checks that the terms
\lstinline{mod2 (S (S (S (S O))))} and \lstinline{O} are convertible.

\subsection{The $\varsigma$-calculus}

Abadi and Cardelli \cite{AbadiCardelli} have introduced a family of calculi
for object-oriented programming. These calculi differ on their typing
discipline and on whether they are purely functional or
imperative. The simplest of them is \textit{Obj}$_1$. It is
object-based---classes are not primitive
constructs but can be defined from objects---and its type system is
similar to that of the simply-typed $\lambda$-calculus.

Types $A$ are records of types $A ::= [l_i : A_i]_{i \in 1..n}$. Labels
are distinct and their order is irrelevant. We translate them
by first sorting the labels and then using association lists

\begin{lstlisting}
label : Type.
type : Type.
typenil : type.
typecons : label -> type -> type -> type.
\end{lstlisting}

\medskip

\(|[l_i : A_i]_{i \in 1\ldots n, l_1 < \ldots < l_n}| = \texttt{typecons}~l_1~|A_1|~(\ldots \texttt{typecons}~l_n~|A_n|~\texttt{typenil} \ldots)\)

\medskip

Objects are records of methods introduced by a special binder
$\varsigma$, binding the object itself inside the method's body. The
primitive operations are method selection and method update:
$$\begin{array}{lcll}
  a, b, \ldots & ::= & x & \emph{(variable)}\\
  & \vert & [l_i=\varsigma(x_i : A) a_i]_{i \in 1 \ldots n} & \emph{(object)}\\
  & \vert & a.l & \emph{(method selection)}\\
  & \vert & a.l \Leftarrow \varsigma(x : A) b & \emph{(method update)}\\
\end{array}$$

We cannot define objects as lists in \textsc{Dedukti}, because a sublist of a
well-typed object is not a well-typed object: the type annotation on
each $\varsigma$ binder refers to the type of the full object.  In
order to define partially constructed objects as lists, we introduce
the notion of preobject.

\begin{lstlisting}
Preobj : type -> type -> Type.
\end{lstlisting}

The first type argument of the type \texttt{Preobj} is the type of the
object that we are constructing. The second argument is the type of
the part we have already constructed, a dependent type
of lists parameterized by the corresponding portion of the type. When
this second part reaches the first one, we get an object, so the type
for objects is
\begin{lstlisting}
def Obj (A : type) := Preobj A A.
\end{lstlisting}

The constructors of preobjects are

\begin{lstlisting}
prenil : A : type -> Preobj A typenil.
precons : A : type ->
          B : type ->
          l : label ->
          C : type ->
          (Obj A -> Obj C) ->
          Preobj A B -> Preobj A (typecons l C B).
\end{lstlisting}

The two primitives on objects, selection and update, are the only
functions that recurse on the list structure of preobjects.  The
update function on preobjects \texttt{preupdate} returns a preobject
of the same type whereas the selection function on preobjects
\texttt{preselect} returns a function defined on \texttt{Obj A}:
\begin{lstlisting}

mem : label -> type -> type -> Type.
athead : l : label -> A : type -> B : type -> mem l A (typecons l A B).
intail : l : label -> A : type -> l' : label -> A' : type
              -> B : type -> mem l A B -> mem l A (typecons l' A' B).

def preselect : A : type ->
                B : type ->
                l : label ->
                C : type ->
                mem l C B ->
                Preobj A B ->
                (Obj A -> Obj C).
def preupdate : A : type ->
                B : type ->
                l : label ->
                C : type ->
                mem l C B ->
                Preobj A B ->
                (Obj A -> Obj C) ->
                Preobj A B.
\end{lstlisting}

Both functions are defined by structural induction on their argument
of type \texttt{mem l C B} where \texttt{mem} is the inductive relation
representing membership of the couple $(\texttt{l}, \texttt{C})$ to a
type \texttt{B}.

From these, selection and update can be defined on objects:
\begin{lstlisting}
def select (A : type) (l : label) (C : type)
           (p : mem l C A) (a : Obj A) : Obj C :=
   preselect A A l C p a a.
def update (A : type) := preupdate A A.
\end{lstlisting}

\begin{definition}[Expression of the $\varsigma$-calculus]
The encoding of $\varsigma$-terms is then defined by
\begin{itemize}
\item $|x| =$ \lst{x},
\item $|[l_i = \varsigma(x : A) a_i]_{i \in 1\ldots n, l_1 < \ldots < l_n}| =$
\lst{precons} $|A|~|[l_i:A_i]_{i\in 2 \ldots n}|$ \lst{l}$_{\tt 1}$~$|A_1|$
\lst{(x : Obj} $|A|$ \lst{=>} $|a_1|$\lst{)}

\hfill
\lst{(... precons} $|A|~|[~]|$ \lst{l}$_{\tt n}$~$|A_n|$
\lst{(x : Obj} $|A|$ \lst{=>} $|a_n|$ \lst{)}~\lst{(prenil}~$|A|$~\lst{)...)},
\item $|a.l_i|  =$ \lst{select} $|A|$ \lst{l}$_{\tt i}$ $|A_i|$ $\tt p_i$ $|a|$,
\item $|a.l_i\Leftarrow\varsigma(x : A) c| =$ \lst{update} $|A|$
\lst{l}$_{\tt i}$ $|A_i|$ $\tt p_i$ $|a|$ \lst{(x : Obj} $|A|$ \lst{=>} $|c|$\lst{)}.
\end{itemize}
where $\tt p_i$ is a proof that $A$ contains $l_i : A_i$.
\end{definition}

\begin{lemma}[Shallow encoding, Theorems 13 and 18 in
\cite{CauderlierDubois2014}]
 This encoding preserves binding, typing, and the operational semantics of the simply-typed $\varsigma$-calculus.
\end{lemma}

This translation of the simply-typed $\varsigma$-calculus in the
$\lambda\Pi$-calculus modulo theories can be extended to subtyping,
see \cite{CauderlierDubois2014}.

\subsection{ML}

The translation of ML types has already been addressed in
Section~\ref{simple-type-theory}. The main features added by the ML
programming language are call by value, recursion and algebraic
datatypes. All these features are present in the $\lambda\Pi$-calculus
modulo theory, but they are only available at toplevel and the notion
of pattern matching in rewrite systems is slightly different. While,
in functional languages, only values are matched and the order in
which patterns are given is usually relevant, rewrite rules can be
triggered on open terms and rewrite systems are assumed confluent.

Making use of tuples if necessary, we assume that every constructor has
arity one. Pattern matching can be faithfully encoded without appealing to
complex compilation techniques thanks to
\emph{destructors}. Destructors generalize over the if-then-else
construct (the destructor associated with the constructor
\texttt{true}) by binding the subterm $t$. Unlike the eliminators of
Section~\ref{inductives}, destructors are specific to a given
constructor, which allows to handle nested patterns.

\begin{definition}
  Let \lst{C} be a datatype constructor of type $\tau$ \lst{->} $\tau'$, the
  destructor associated with \lst{C} is the symbol
  $\mbox{\lst{destr}}_{\mbox{\lst{C}}}$, of type
  \lst{R : type -> (eps} $\tau'$ \lst{-> (eps} $\tau$ \lst{-> eps R) -> eps R} \lst{-> eps R)}, defined by the following rewrite rules:
$$
\begin{array}{c@{~}c@{~}c@{~}c@{~}ccll}
  \mbox{\lst{destr}}_{\mbox{\lst{C}}} & \mbox{\lst{R}} & \mbox{\lst{(C t)}}  & \mbox{\lst{f}} & \mbox{\lst{d}} & \mbox{\lst{-->}} & \mbox{\lst{f t}} & \\
  \mbox{\lst{destr}}_{\mbox{\lst{C}}} & \mbox{\lst{R}} & \mbox{\lst{(C' t')}}& \mbox{\lst{f}} & \mbox{\lst{d}} & \mbox{\lst{-->}} & \mbox{\lst{d}} &
  \mbox{(for all other constructors \lst{C'} for type $\tau'$)} 
\end{array}
$$
\end{definition}

\begin{lemma}[Theorem 1 in \cite{CauderlierDubois2016}]
  ML pattern matching can be expressed by destructors in a
  semantics-preserving way.
\end{lemma}

Recursion can also be encoded without generating obvious
non-termination on open terms; we introduce a global symbol \lst{@},
of type
\lst{A : type -> B: type -> ((eps A -> eps B)} \lst{-> eps A -> eps B)},
defined for each constructor \lst{C} of type $\tau$ \lst{->} $\tau'$ by the
following rewrite rule
\begin{center}
\lst{@} $\tau'$ \lst{R f (C t) --> f (C t)}
\end{center}

This symbol freezes evaluation until the argument
starts with a constructor. It is inserted in the right-hand side
of all recursive definitions to freeze recursive calls.

\begin{lemma}
  If a ML term $t$ evaluates to a ML value $v$, then its translation
  $|t|$ reduces to $|v|$.
\end{lemma}

\subsection{\textsc{FoCaLiZe}}

Our main motivation for translating programming languages to the
$\lambda\Pi$-calculus modulo theory is to check program
certificates.  The \textsc{FoCaLiZe}
system \cite{FoCaLiZe} is an environment for
the development of certified programs; it uses ML as an implementation
language, classical predicate logic as a specification language, and
provides static object-oriented features to ease
modularity. Moreover, it delegates most proofs to \textsc{Zenon
  Modulo}, which provides \textsc{Dedukti} proofs, as seen in
Section~\ref{zenon}. Thus, it is a good candidate for an encoding to
\textsc{Dedukti}.

Let us consider again the function \texttt{mod2}. It can be defined in
  \textsc{FoCaLiZe} by
\begin{verbatim}
type nat = | Zero | Succ (nat);;

let rec mod2 (n) =
  match n with
  | Zero -> Zero
  | Succ(m) ->
     match m with
       | Zero -> Succ(Zero)
       | Succ(k) -> mod2(k)
;;
\end{verbatim}

  A specification for \texttt{mod2} would typically include a theorem
  stating that \texttt{mod2} returns \texttt{Zero} on even numbers,
  where by ``even number'' we mean a number which is the double of
  another.  This can be formulated from the function \texttt{twice}
  returning the double of a number:
\begin{verbatim}
let rec twice (n) =
  match n with
  | Zero -> Zero
  | Succ(m) -> Succ(Succ(twice(m)))
;;
\end{verbatim}

  Hence, the statement of our theorem would be \lst{all n:nat,
    mod2(twice(n)) = Zero}.  This theorem is proved by induction over
  natural numbers. The two inductive steps can be processed by
  \textsc{Zenon Modulo}.

\begin{verbatim}
theorem mod2_twice_Zero : mod2(twice(Zero)) = Zero
proof = by type nat definition of mod2, twice;;

theorem twice_Succ : all n : nat, twice(Succ(n)) = Succ(Succ(twice(n)))
proof = by type nat definition of twice;;

theorem mod2_SuccSucc : all n : nat, mod2(Succ(Succ(n))) = mod2(n)
proof = by type nat definition of mod2;;

theorem mod2_twice_Succ :
  all n : nat,
    mod2(twice(n)) = Zero ->
    mod2(twice(Succ(n))) = Zero
proof = by type nat property mod2_SuccSucc, twice_Succ;;
\end{verbatim}

As we can see from this example, \textsc{FoCaLiZe}
proofs are very succinct; they are mere a list of hints, used to
compute a  first-order problem, that is given to \textsc{Zenon
  Modulo}.  In order to conclude the proof of the theorem, we need
an induction principle. It can be  axiomatized in \textsc{FoCaLiZe}

\begin{verbatim}
theorem nat_induction :
  all p : (nat -> bool),
    p(Zero) ->
    (all n : nat, p(n) -> p(Succ(n))) ->
    all n : nat, p(n)
proof = assumed;;
\end{verbatim}
but this is not a first-order statement and \textsc{Zenon Modulo} is
not able to instantiate it. However \textsc{FoCaLiZe} allows to write
the instantiation directly in Dedukti
\begin{verbatim}
let mod2_twice_n_is_Zero (n : nat) : bool = (mod2(twice(n)) = Zero);;

theorem mod2_twice :
  all n : nat, mod2(twice(n)) = Zero
proof =
  dedukti proof
  {* nat_induction mod2_twice_n_is_Zero mod2_twice_Zero mod2_twice_Succ. *};;
\end{verbatim}

\textsc{FoCaLiZe} version 0.9.2 features \textsc{Dedukti} code generation.
More than 98\% of \textsc{FoCaLiZe} standard library is checked in
\textsc{Dedukti}.  The size of the gzipped \textsc{Dedukti}
translation weighs 1.89 MB.

\section{The Calculus of inductive constructions with universes}
\label{sec:cicw}

\subsection{Pure type systems}
\label{pure-type-systems}

Pure type systems \cite{Barendregt1992} are a general class of typed
$\lambda$-calculi that include many important logical systems based on
the propositions-as-types principle such as the Calculus of
constructions. We can express functional Pure type systems in the
$\lambda\Pi$-calculus modulo theory \cite{CousineauDowek}
using ideas similar to the
expression of Simple type
 theory, see Section~\ref{simple-type-theory}, and to the mechanisms of
Tarski-style
universes
in Intuitionistic type theory \cite{MartinLof}. We
represent each type $A$ of sort $s$ as a term $|A|$, of type
$U_s$, and each term $M$ of type $A$ as a term $|M|$, of type
\lst{eps}$_s~|A|$ \cite{CousineauDowek}.

Consider a Pure type system specification
$(\mathcal S, \mathcal A, \mathcal R)$. For each sort $s \in \mathcal
S$, we declare the type $U_s$ whose inhabitants represent the types of
sort $s$, and the dependent type $\mbox{\lst{eps}}_s$  whose inhabitants
represent the terms of the types of sort $s$. For conciseness, in this
Section we use \lst{e} instead of \lst{eps}.

\medskip

\noindent
\lst{U_s : Type.}

\noindent
\lst{def e_s : U_s -> Type.}

\medskip

\noindent
For each axiom $\langle s_1, s_2 \rangle \in \mathcal A$, we declare the
variable
\lst{u}$_{s_1}$ which is the term that represents the sort $s_1$ in $s_2$,
and we give its interpretation as a type using a rewrite rule.

\medskip

\noindent
\lst{u_s1 : U_s2.}

\noindent
\lst{[ ] e_s2 u_s1 --> U_s1.}

\medskip

\noindent
For each rule $\langle s_1, s_2, s_3 \rangle
\in \mathcal R$, we declare the variable
$\pi_{s_1,s_2}$ which is the term that represents dependent products such as $\Pi x : A~B$ in $s_3$, and we give its interpretation as a type using a rewrite rule.

\noindent
\lst{pi_s1s2 : a : U_s1 -> b : (e_s1 a -> U_s2) -> U_s3.}

\noindent
\lst{[a, b] e_s3 (pi_s1s2 a b) --> x : e_s1 a -> e_s2 (b x).}

\begin{definition}
The terms of the Pure type system are expressed as
\begin{itemize}
\item $|x| =$ \lst{x},
\item $|s| =$ \lst{u_s},
\item $|M~N| = |M|~|N|$,
\item $|\lambda x:A~M| =$ \lst{x :} $\|A\|$ \lst{=>} $|M|$,
\item $|\Pi x:A~B| =$ \lst{pi_s1s2} $|A|$ \lst{(x:} $\|A\|$
\lst{=>} $|B|$\lst{)}
(where $\Gamma \vdash A : s_1$ and $\Gamma, x:A \vdash B : s_2$),
\end{itemize}

\begin{itemize}
\item $\|A\| =$  \lst{e_s} $|A|$ (when $\Gamma \vdash A : s$),
\item $\|s\| =$ \lst{U_s} (when $\nexists s'$ such that $\Gamma \vdash s : s'$),
\end{itemize}

\begin{itemize}
\item $\|[~]\| = [~]$,
\item $\|\Gamma, x:A\|  =  \|\Gamma\|,$ \lst{x:}$\|A\|$.
\end{itemize}
\end{definition}

\begin{lemma}[Preservation of computation, Proposition 1 in
\cite{CousineauDowek}]
If $\Gamma \vdash M : A$ and $M \longrightarrow_\beta M'$, then there
exists $N'$ such that $|M| \longrightarrow_\beta N' \equiv_{\beta\Sigma} |M'|$.
\end{lemma}


\begin{lemma}[Preservation of typing, Proposition 3 in \cite{CousineauDowek}]
If $\Gamma \vdash M : A$, then $\Sigma, \|\Gamma\| \vdash |M| : \|A\|$.
\end{lemma}

It is also conservative---that is, adequate---with respect to the
original system.

\begin{lemma}[Conservativity, adequacy, \cite{Assaf2015}]
If $\Sigma, \|\Gamma\| \vdash M' : \|A\|$, then there exists $M$ such that $\Gamma \vdash M : A$ and $|M| \equiv_{\beta\eta\Sigma} M'$.
\end{lemma}

Together, these results show that provability in the Pure type system
is equivalent to provability in the embedding.
\begin{theorem}[\cite{Assaf2015}] There exists $M$, such that $\Gamma
  \vdash M : A$ if, and only if there exists $M$, such that $\Sigma,
  \|\Gamma\| \vdash M : \|A\|$.
\end{theorem}

As an example, the Calculus of constructions can be defined as the Pure type system
$$
\begin{array}{lll}
\mathcal S & ::= & \{Type, Kind\} \\
\mathcal A & ::= & \{(Type, Kind)\} \\
\mathcal R & ::= & \{(Type,Type,Type), (Type,Kind,Kind), \\
           &   & \phantom{\{}(Kind, Type, Type), (Kind, Kind, Kind)\} \\
\end{array}
$$
and it is expressed in \textsc{Dedukti} as
\begin{lstlisting}
U_Type : Type.
U_Kind : Type.

def e_Type : U_Type -> Type.
def e_Kind : U_Kind -> Type.

u_Type : U_Kind.

pi_TypeType : a : U_Type -> b : (e_Type a -> U_Type) -> U_Type.
pi_TypeKind : a : U_Type -> b : (e_Type a -> U_Kind) -> U_Kind.
pi_KindType : a : U_Kind -> b : (e_Kind a -> U_Type) -> U_Type.
pi_KindKind : a : U_Kind -> b : (e_Kind a -> U_Kind) -> U_Kind.

[ ] e_Kind u_Type --> U_Type.

[a, b]
  e_Type (pi_TypeType a b) --> x : e_Type a -> e_Type (b x).
[a, b]
  e_Kind (pi_TypeKind a b) --> x : e_Type a -> e_Kind (b x).
[a, b]
  e_Type (pi_KindType a b) --> x : e_Kind a -> e_Type (b x).
[a, b]
  e_Kind (pi_KindKind a b) --> x : e_Kind a -> e_Kind (b x).
\end{lstlisting}
Consider the context $\Gamma = a : Type, b : Type, x : a, f : \Pi p : (a \to Type)~(p~x \to b)$, in which the term $f~(\lambda y:a~a)~x$ is a proof of $b$. This proof is represented in \textsc{Dedukti} as
\begin{lstlisting}
a : U_Type.
b : U_Type.
x : e_Type a.
f : (p : (e_Type a -> U_Type) -> e_Type (p x) -> e_Type b).

def example : e_Type b := f (y : e_Type a => a) x.
\end{lstlisting}

\subsection{Inductive types}
\label{inductives}

The Calculus of inductive constructions is an extension of the
Calculus of constructions with inductive types.

The straightforward way to encode inductive types in the
$\lambda\Pi$-calculus modulo theory is to use constructors and a
primitive recursion operator as in G\"odel's system T \cite{Godel58}.
For example, for polymorphic lists, we declare a new
variable \lst{list}, such that \lst{list A} is the
type of lists of elements of \lst{A} and two variables
\lst{nil} and \lst{cons} for the constructors:
\begin{lstlisting}
list : U_Type -> U_Type.
nil : A : U_Type -> e_Type (list A).
cons : A : U_Type -> e_Type A -> e_Type (list A) -> e_Type (list A).
\end{lstlisting}

We still need an elimination scheme to define functions by recursion on lists and to prove theorems by induction on lists. To this end, we declare a new
variable \lst{elim_list} for the primitive recursion operator:
\begin{lstlisting}
List : U_Type -> Type.
[A] e_Type (list A) --> List A.

def elim_list : A : U_Type -> P : (e_Type (list A) -> U_Type)
  -> e_Type (P (nil A)) -> (x : e_Type A -> l2 : e_Type (list A)
  -> e_Type (P l2) -> e_Type (P (cons A x l2))) -> l : e_Type (list A)
  -> e_Type (P l).
\end{lstlisting}

The operator is parameterized by two types: the type \lst{A} of the
elements of the list we are eliminating and the return type \lst{P},
which is the type of the object we are constructing by eliminating this list.
It is a dependent type because it can depend on the list. The type of
\lst{elim_list} expresses that, if we provide a base case of type
\lst{P nil}, and a way to construct, for any element
\lst{x:A} and list \lst{l2:(list A)} and object of type
\lst{P l2}, an object of type \lst{P (cons A x l2)}, then we
can construct an object of type \lst{P l} for any list \lst{l}.
Finally, we express the computational behavior of the
eliminator, that reduces when applied to a constructor.
\begin{lstlisting}
[A, P, case_nil, case_cons]
  elim_list _ P case_nil case_cons (nil A) -->
  case_nil.
[A, P, case_nil, case_cons, x, l]
  elim_list _ P case_nil case_cons (cons A x l) -->
  case_cons x l (elim_list A P case_nil case_cons l).
\end{lstlisting}

Recursive functions and inductive proofs can then be expressed using
this elimination operator. For instance, the append function is
\begin{lstlisting}
def append : A : U_Type -> List A -> List A -> List A :=
  A : U_Type => l1 : List A => l2 : List A =>
  elim_list A (__ => list A) l2 (x => l3 => l3l2 => cons A x l3l2) l1.
\end{lstlisting}
We can extend this technique to other inductive types, and we can also
adapt it to represent functions defined by pattern matching and
guarded recursion---as opposed to ML general recursion---, like those of the
\textsc{Coq} system \cite{BoespflugBurel2012}.
Note that to formalize the full Calculus of
inductive constructions,  we need an infinite number of rules. As
noticed in Section \ref{sec:infinite}, each proof uses only a finite
number of rules: those of the inductive types used in this proof.

\subsection{Universes}
\label{universes}

Universes are the types of types in Intuitionistic type theory
\cite{MartinLof}. They are similar to sorts in Pure type
systems. To fully encode the universes of Intuitionistic type theory,
we need to take into account two more features. First, there is an
infinite hierarchy of universes
\[
U_0 : U_1 : U_2 : \cdots
\]
Second, the hierarchy is cumulative
\[
U_0 \subseteq U_1 \subseteq U_2 : \cdots
\]
which is expressed by the typing rule
$$\irule{\Gamma \vdash M : U_i}{\Gamma \vdash M : U_{i+1}}{}$$
To deal with the infinite universe hierarchy, instead of declaring a
different variable for each universe, we index the variable $U,
\mbox{\lst{eps}}, u, \pi$ by natural numbers.
\begin{lstlisting}
nat : Type.
0 : nat.
S : nat -> nat.

U : nat -> Type.
def eps : i : nat -> U i -> Type.

u : i : nat -> U (S i).
[i] eps _ (u i) --> U i.

def pi : i : nat -> a : U i -> b : (eps i a -> U i) -> U i.
[i, a, b] eps _ (pi i a b) --> x : eps i a -> eps i (b x).
\end{lstlisting}

To express cumulativity, we cannot just identify the universe $U_i$
with the universe $U_{i+1}$, because not all terms of type $U_{i+1}$
have type $U_i$. In particular $U_{i}$ itself does not.  We therefore
rely on explicit cast operators $\uparrow_i$ that lift types from a
universe $U_i$ to a higher universe $U_{i+1}$.
\begin{lstlisting}
lift : i : nat -> U i -> U (S i).
[i, a] eps _ (lift i a) --> eps i a.
\end{lstlisting}

A side effect of introducing explicit casts is that some types can
have multiple representations as terms. For example, if $\Gamma \vdash
A : U_i$ and $\Gamma, x:A \vdash B : U_i$ then the type $\Pi x:A~B$
has two different representations as a term in the universe $U_{i+1}$:
\lst{lift i (pi i} $|A|$ \lst{(x =>} $|B|$\lst{)} and
\lst{pi (S i) (lift i} $|A|$ \lst{) (x => lift i} $|B|$\lst{)}.
This
multiplicity can break the preservation of typing \cite{Assaf2014}, so
we add the following rewrite rule to ensure that types have a unique
representation as terms:
\begin{lstlisting}
[i, a, b] pi _ (lift i a) (x => lift {i} (b x)) -->
          lift i (pi i a (x => b x)).
\end{lstlisting}

These techniques can also be adapted to express the universes of the
Calculus of inductive constructions, which in particular include an
impredicative universe $Prop$ \cite{Assaf2014}.

\subsection{\textsc{Matita} proofs}

There are not many libraries of proofs expressed in the pure Calculus of
constructions with universes. Indeed, the \textsc{Coq} library is
expressed in a Calculus of constructions with universes, modules and
universe polymorphism. Those two latter features have not yet been
expressed in \textsc{Dedukti}, so the \textsc{Coq} library cannot be
checked directly. Preliminary results in this direction are
presented in \cite{Jouannaud}.

A closely related library comes from \textsc{Matita}. It is expressed
in a Calculus of constructions with universes, and proof irrelevance.
Even if proof irrelevance has not yet been expressed in
\textsc{Dedukti}, the system \textsc{Krajono} is able to translate all
the files which come with the \textsc{Matita} tarball and which do not
explicitly use proof irrelevance. Tests have been conducted on the
\texttt{arithmetic} library of Matita, giving a successfully checked
\textsc{Dedukti} library of 1.11 MB, when gzipped.

\section{Contributions}

The results presented in this paper build upon the work of many people.

The $\lambda \Pi$-calculus modulo theory appeared in a joint paper of
Denis Cousineau and Gilles Dowek in 2007 \cite{CousineauDowek},
together with an expression of Pure type systems in it, and a partial
correctness proof. The framework was further refined in \cite{RSai13}
and then in \cite{Saillard15}. In 2010, during his undergrad internship
\cite{Dorra}, Alexis Dorra showed that constructive predicate logic
also could be expressed in the $\lambda \Pi$-calculus modulo theory.

The first implementation of \textsc{Dedukti} was designed by Mathieu
Boespflug \cite{Boespflug} between 2008 and 2011, during his thesis
supervised by Gilles Dowek. A second implementation was designed in 2012
by Quentin Carbonneaux during his Master thesis \cite{Carbonneaux},
supervised by Olivier Hermant and Mathieu Boespflug.  Finally, the
current implementation has been designed by Ronan Saillard between
2012 and 2015 during his thesis \cite{Saillard15}, supervised by Pierre
Jouvelot and Olivier Hermant.

A first expression of the Calculus of Inductive Constructions was
proposed by Mathieu Boespflug and Guillaume Burel in 2011
\cite{BoespflugBurel2012}.  An expression of Simple type theory and of
the Calculus of Inductive Constructions was then proposed by Ali Assaf
between 2012 and 2015 during his thesis \cite{AssafThese} advised by
Gilles Dowek and Guillaume Burel. This led to the translations of the
proofs of \textsc{HOL Light} and \textsc{Matita} in
\textsc{Dedukti}. A graphical front-end for the translation of the proofs of
\textsc{Holide} has been proposed by Shuai Wang during his Master thesis
in 2015 \cite{Wang}.
Guillaume Burel \cite{Burel13} showed in 2013 that
\textsc{iProverModulo} could produce \textsc{Dedukti} proofs. During
his thesis, started in 2013, advised by David Delahaye, Pierre
Halmagrand showed that \textsc{Zenon Modulo} also could produce
\textsc{Dedukti} proofs. Later Pierre Halmagrand, Fr\'ed\'eric Gilbert,
and Rapha\"el Cauderlier, showed how classical proofs could be
expressed in \textsc{Dedukti} without any axiom, introducing the classical
connectives and quantifiers. During his thesis,
started in 2013, advised by Catherine Dubois, Rapha\"el Cauderlier
opened a new direction, with the expression of programs and properties
of programs, in particular those expressed in \textsc{FoCaLiZe}.

\section{Conclusion and future work}

The encodings and the benchmarks discussed in this paper give an
overview of the versatility of \textsc{Dedukti} to express a wide
range of theories. They also demonstrate that this tool scales up well
to very large libraries, recall the five libraries discussed in this paper:
\begin{tabbing}
\= aaaaaaaaa \= The Zenon modulo Set Theory Library aaa \= 595 MB\kill
\> \> The iProverModulo TPTP library \> 38.1 MB\\
\> \> The Zenon modulo Set Theory Library \> 595 MB\\
\> \> The Focalide library \> 1.89 MB\\
\> \> The Holide library \> 21.5 MB\\
\> \> The Matita arithmetic library \> 1.11 MB
\end{tabbing}
Like the \textsc{Dedukti} system, all these translators and libraries
are available on the \textsc{Dedukti} web site.

Future work include expressing more proofs in \textsc{Dedukti}, in
particular proofs coming from \textsc{Coq}, \textsc{PVS} and from SMT
solvers and build a larger proof library.

It also includes reverse engineering of proofs. Although
\textsc{HOL Light} is a classical system, many proofs of the
\textsc{HOL Light} library happen to be constructive.  Many proofs
in the \textsc{Matita} library do not require the full power of the
Calculus of inductive constructions with universes and can be
expressed in weaker theories.

This reverse engineering of proofs is a first step towards the
possibility to use \textsc{Dedukti} to develop complex proofs by
combining lemmas developed in several systems.  A first investigation
in this direction is presented in \cite{AssafCauderlier2015},
but still requires to be generalized and automated.

More generally, we hope, with this project, to contribute to the shift
of the general question ``What is a good system to express
mathematics?''  to the more specific questions ``Which definitions,
axioms and rewrite rules are needed to prove which theorem?''

\section*{Acknowledgements}

Many thanks to
Jasmin Blanchette,
Fr\'ed\'eric Blanqui,
Cezary Kaliszyk,
Claude Kircher, and Jean-Pierre Jouannaud for very helpful comments on
a previous version of this paper.

\bibliographystyle{plain}
\bibliography{expressing}

\end{document}